\documentclass[11pt]{article}
\usepackage{amsmath,amssymb,color,graphics,epsfig}

\usepackage{amsfonts}

\newcommand{\hoch}[1]{$\, ^{#1}$}

\makeatletter
\@addtoreset{equation}{section}
\makeatother

\newcommand{\be}{\begin{equation}}
\newcommand{\ee}{\end{equation}}
\newcommand{\bea}{\setlength\arraycolsep{2pt} \begin{eqnarray}}
\newcommand{\eea}{\end{eqnarray}}
\newcommand{\nn}{\nonumber}

\def\ft#1#2{{\textstyle{\frac{\scriptstyle #1}{\scriptstyle #2} } }}
\def\fft#1#2{{\frac{#1}{#2}}}

\def\0{{\sst{(0)}}}
\def\1{{\sst{(1)}}}
\def\2{{\sst{(2)}}}
\def\3{{\sst{(3)}}}
\def\4{{\sst{(4)}}}
\def\5{{\sst{(5)}}}
\def\6{{\sst{(6)}}}
\def\7{{\sst{(7)}}}
\def\8{{\sst{(8)}}}
\def\sst#1{{\scriptscriptstyle #1}}
\def\oneone{\rlap 1\mkern4mu{\rm l}}

\def\del{{\partial}}

\def\im{{{\rm i}}}

\def\th{{{f}}}
\def\cR{{{\cal R}}}

\begin{document}

\begin{flushright}
MIFPA-14-05

\end{flushright}

\vspace{25pt}
\begin{center}
{\large {\bf Thermodynamics of Einstein-Proca AdS Black Holes}}

\vspace{10pt}
Hai-Shan Liu\hoch{1}, H. L\"u\hoch{2} and C.N. Pope\hoch{3,4}

\vspace{10pt}

\hoch{1} {\it Institute for Advanced Physics \& Mathematics,\\
Zhejiang University of Technology, Hangzhou 310023, China}

\vspace{10pt}

\hoch{2}{\it Department of Physics, Beijing Normal University,
Beijing 100875, China}

\vspace{10pt}

\hoch{3} {\it George P. \& Cynthia Woods Mitchell  Institute
for Fundamental Physics and Astronomy,\\
Texas A\&M University, College Station, TX 77843, USA}

\vspace{10pt}

\hoch{4}{\it DAMTP, Centre for Mathematical Sciences,
 Cambridge University,\\  Wilberforce Road, Cambridge CB3 OWA, UK}

\vspace{40pt}

\underline{ABSTRACT}
\end{center}

   We study static spherically-symmetric solutions of the Einstein-Proca
equations in the presence of a negative cosmological constant.  We
show that the theory admits solutions describing both black holes and also
solitons in an asymptotically AdS background.  Interesting subtleties
can arise in the computation of the mass of the solutions and also in
the derivation of the first law of thermodynamics.  We make use of
holographic renormalisation in order to calculate the mass, even in cases where
the solutions have a rather slow approach to the asymptotic AdS geometry.
By using the procedure developed by Wald, we derive the first law of
thermodynamics for the black hole and soliton solutions.  This includes a
non-trivial contribution associated with the Proca ``charge.''  The
solutions cannot be found analytically, and so we make use of numerical
integration techniques to demonstrate their existence.

\thispagestyle{empty}

\pagebreak

\tableofcontents
\addtocontents{toc}{\protect\setcounter{tocdepth}{2}}



\section{Introduction}

   There has been a resurgence of interest recently in constructing
black holes in a variety of theories that, in one way or another, are more
general than those typically considered heretofore.
One particular aspect that is now
receiving considerable attention is the study of black holes in theories
admitting ant-de Sitter (AdS) rather than Minkowski backgrounds, since asymptotically AdS solutions play a central role in the AdS/CFT correspondence
\cite{mald,gkp,wit}.

   Higher-derivative theories of gravity provide a fertile ground for
constructing asymptotically AdS solutions, but they typically suffer from
the drawback that the concomitant massive spin-2 modes have the wrong
sign for their kinetic terms in a linearised analysis around an AdS background,
and thus the theories are, in general, intrinsically plagued by ghosts.
Nevertheless, as
a framework for purely classical investigations, they can provide
interesting starting points for the study of black-hole solutions and
their dynamics.  For example, in some recent work on the existence of
black hole solutions in Einstein-Weyl gravity with a cosmological constant,
it was found through numerical studies that asymptotically-AdS black holes
can arise whenever the mass-squared ${\bf m}_2^2$ of the massive spin-2 mode is
negative \cite{lupapovp}.  In general one may expect a negative mass-squared
to exhibit another undesirable feature, namely the occurrence of tachyonic
run-away instabilities that grow exponentially as a function of time.
However, in anti-de Sitter backgrounds there is a ``window'' of negative
mass-squared values ${\bf m}_{2\sst{{BF}}}^2 \le {\bf m}_2^2 <0$,
where the negative mass-squared ${\bf m}_{2\sst{{BF}}}^2 $ is the
limiting value of the
so-called Breitenlohner-Freedman bound \cite{BF1,BF2},
above  which the linearised
modes in the AdS background still have oscillatory rather than
real exponential time dependence, and thus the  run-away behaviour is avoided.

   In this paper, we shall study a different type of theory where
again certain modes of non-tachyonic negative mass-squared play a central
role, but now in the framework of a more conventional two-derivative
theory that has no accompanying ghost problems.  Specifically, we shall
focus on the $n$-dimensional
Einstein-Proca theory of a massive spin-1 field coupled
to Einstein gravity, in the presence also of a (negative) cosmological
constant.  This theory exhibits many features that are similar to those of a
higher-derivative theory of gravity, with the massive Proca field
now playing the r\^ole of the massive spin-2 mode.
The solutions we study approach anti-de Sitter spacetime at
large distance, with $R_{\mu\nu} \sim -(n-1)\ell^{-2}\, g_{\mu\nu}$ as
the radius tends to infinity.  We find two distinct kinds of
spherically-symmetric static solutions,
namely asymptotically AdS black holes, and smooth asymptotically AdS
solitons.  The metrics for both classes of solution take the form
\be
ds^2 = -h(r)\, dt^2 + \fft{dr^2}{f(r)} + r^2\, d\Omega_{n-2}^2\,.
\label{nmetric}
\ee

   It is useful first to consider the situation where the mass is
small and the Proca field is
weak, so that its back-reaction on the spacetime geometry can be neglected.
In this limit, the large-distance behaviour of the Proca potential
$A=\psi(r)\, dt$ approaches that of a massive vector in AdS, which
takes the form
\be
\psi(r)\rightarrow
  \fft1{r^{(n-3-\sigma)/2}}\, \sum_{p=0}^\infty \fft{b_p}{r^{2p}}
 + \fft1{r^{(n-3+\sigma)/2}}\, \sum_{p=0}^\infty \fft{\tilde b_p}{r^{2p}} \,,
\label{larger}
\ee
where
\be
\sigma= \sqrt{4 \tilde m^2\, \ell^2 + (n-3)^2}\,,\label{sigmadef}
\ee
and $\tilde m$ is the mass of the Proca field.  The Breitenlohner-Freedman
window of non-tachyonic negative mass-squared values lies in the range
${\bf m}_{\sst{BF}}^2 \le \tilde m^2 <0$, with
\be
{\bf m}_{\sst{BF}}^2 = -\ft14\ell^{-2}\, (n-3)^2\,,\label{BFbound}
\ee
thus ensuring that $\sigma$ remains real.
  The sums in (\ref{larger})
can actually be expressed in closed forms as hypergeometric functions
(see appendix A).

   It follows from the equations of motion that
the back-reaction of the Proca field on the geometry will first appear
in the metric functions $h$ and $f$ at order $1/r^{n-3-\sigma}$, together
with associated higher inverse powers of $r$.  The back-reacted geometry will
in turn modify the Proca solution, with the onset beginning at order
$1/r^{(3n-5-3\sigma)/2}$.  Thus at large $r$ the Proca
and metric functions will include terms of the general form
\bea
\psi&=& \fft{q_1}{r^{(n-3-\sigma)/2}} +
    \fft{q_2}{r^{(n-3+\sigma)/2}} + \fft{a_1 \, q_1^3}{r^{(3n-5-3\sigma)/2}}
   +\cdots\,,\nn\\
h &=& r^2\, \ell^{-2} + 1 + \fft{m_2}{r^{n-3}} +
  \fft{a_2\, q_1^2}{r^{n-3-\sigma}}\cdots\,,\nn\\
f &=& r^2\, \ell^{-2} + 1 + \fft{n_2}{r^{n-3}} +
  \fft{a_3\, q_1^2}{r^{n-3-\sigma}}\cdots\,. \label{asymptotics}
\eea
The terms with coefficients $m_2$ and $n_2$ in $h$ and $f$ are associated
with the mass of the black hole or soliton.

    The ellipses denote all the remaining terms in the large-$r$ expansions.
These will include the direct ``descendants'' of the $q_1/r^{(n-3-\sigma)/2}$
and $q_2/r^{(n-3+\sigma)/2}$ terms, at orders $1/r^{(n-3-\sigma)/2+2p}$
and $1/r^{(n-3+\sigma)/2+2p}$ for all integers $p\ge1$, as in (\ref{larger}),
and also higher-order back-reaction terms and descendants of these.
Depending on the value of the index $\sigma$, defined by (\ref{sigmadef}),
some of these remaining terms may intermingle with orders already
displayed in (\ref{asymptotics}), or they may all be at higher orders
than the displayed terms.   The discussion of the asymptotic forms of the
solutions can therefore become quite involved in general.  It will be
convenient for some of our calculations to focus the cases where $\sigma$
is sufficiently small that the displayed terms in (\ref{asymptotics}) are
in fact the leading order ones, and all the terms represented by the
ellipses are of higher order than the displayed ones.  This will
certainly be the case, in all dimensions $n\ge 4$, if we choose
\be
\sigma\le 1\,.\label{sigmale1}
\ee
This allows us to investigate the asymptotic structure of the solutions
systematically for a non-trivial range of Proca mass values corresponding to
$0\le\sigma\le 1$, near to the Breitenlohner-Freedman bound.

   In fact the special case when $\sigma=1$ is particularly nice,
since then the characterisation of inverse powers of $r$ that arise in
the asymptotic expansions of the Proca and metric functions is
very simple.  In this special case the leading-order term in
the expansion of $\psi(r)$ is $1/r^{(n-4)/2}$, with each successive term
having one extra power of $1/r$.  The terms in the metric functions
occur always at integer powers of $1/r$:
\bea
\psi(r) &=&\fft1{r^{(n-4)/2}}\, \sum_{p=0}^\infty \fft{q_{p+1}}{r^p}\,,\cr
h(r) &=& r^2\, \ell^{-2} + 1 + \fft1{r^{n-4}}\,
  \sum_{p=0}^\infty \fft{m_{p+1}}{r^p}\,,\qquad
f(r) = r^2\, \ell^{-2} + 1 + \fft1{r^{n-4}}\,
  \sum_{p=0}^\infty \fft{n_{p+1}}{r^p}\,.\label{sig1exp}
\eea
Accordingly, we first study the asymptotic expansions
in this special case with $\sigma=1$, where it is rather straightforward to
see how one can systematically solve, order by order in powers of
$1/r$, for all the coefficients in terms of $q_1$, $q_2$ and $m_2$.
We then look at the leading orders in the expansions for the general
case $\sigma\le 1$, sufficient for our subsequent purposes of
computing the physical mass and deriving the first law of thermodynamics.
We also study some isolated examples
with $\sigma>1$, showing that even though the structure of the inverse
powers of $r$ in the asymptotic expansions can then be rather involved,
the solutions can still be found in these cases.

   We then turn to the problem of calculating
the physical mass of the solutions in terms of the free
adjustable parameters of the asymptotic expansions, which can be taken to
be $q_1$, $q_2$ and $m_2$.
The calculation turns out to be somewhat non-trivial, because of the
presence of the terms in the metric functions $h$ and $f$ at order
$1/r^{n-3-\sigma}$, which represents a slower fall-off than the
usual $1/r^{n-3}$ of a normal mass term in $n$ dimensions.  Indeed, we
find that a naive calculation using the prescription of
Ashtekar, Magnon and Das (AMD) \cite{ashmag,ashdas}, in which a certain
electric component of the Weyl tensor in a conformally-related metric is
integrated over the boundary $(n-2)$-sphere, leads to a divergent result.
Presumably a more careful analysis, taking into account boundary contributions
that can normally be neglected, may give rise to a finite and meaningful
result.  In the present paper we opt instead for calculating the mass
using the method of the holographic stress tensor, and thereby obtain a
well-defined finite result.

  Having obtained an expression for the physical mass in terms of the
parameters characterising the asymptotic form of the solution we then
study the first law of thermodynamics, using the methods developed by
Wald.  We find that the Proca field makes a non-trivial contribution
in the first law, which, for the static solutions we consider in this
paper, will now take the form
\be
dM = T dS -\fft{\sigma\, \omega_{n-2}}{4\pi}\, q_1\, dq_2\,,
\label{firstlaw0}
\ee
where $\omega_{n-2}$ is the volume of the unit $(n-2)$-sphere.  The
way in which the Proca field contributes to the first law is analogous
to a phenomenon that has been encountered recently when studying the
thermodynamics of dyonic black holes in certain gauged supergravities,
where parameters characterising the asymptotic behaviour of a
scalar field enter in the first law, thus indicating the presence of a
scalar ``charge'' or hair \cite{gaugedyon}.  (See \cite{Liu:2013gja} for
higher-dimensional generalisations.)  In the present case, one
can think of the asymptotic parameters $q_2$ and $q_1$ as being like
a thermodynamic conjugate charge and potential pair characterising
Proca ``hair.''

   The equations of motion for the Einstein-Proca theory appear not to be
exactly solvable even in the static spherically-symmetric situation that
we study in this paper.  We therefore turn to a numerical analysis in
order to establish more explicitly the nature of the solutions to the
theory.  We do this by first developing small-distance series expansions
for the metric and Proca functions, and then using these to set initial
data for a numerical integration out to large distance.  We find two
different kinds of regular short-distance behaviour.  In one type, we integrate
out from a black-hole horizon located at some radius $r_0>0$ at which
the Proca and metric functions all vanish.  In the
other type, we start from a smooth coordinate origin at $r=0$, where
the Proca and metric functions all begin with non-vanishing constant values.
In each of these types of solution, the numerical integration indicates
that the fields stably approach the expected asymptotic forms we discussed
above, thus lending confidence to the idea that such well-behaved
black hole and soliton solutions do indeed exist.  By matching the
numerical solutions to the expansions (\ref{asymptotics})
in the asymptotic region, we can relate the mass and Proca charge to
black-hole area and surface gravity, and thereby obtain numerical
confirmation of the first law (\ref{firstlaw0}).

   The organisation of the paper is as follows.  In section 2 we present
the Lagrangian and equations of motion for the Einstein-Proca theory, and
the consequent ordinary differential equations that are satisfied by the
Proca potential and the metric functions in the static spherically-symmetric
ansatz.  We then examine the asymptotic forms of the solutions that arise
when both of the parameters characterising the Proca field are turned on.
In section 3, we show how the holographic renormalisation procedure
may be used to calculate the mass of the solutions.  In section 4, we
use the formalism developed by Wald in order to derive the first law
of thermodynamics for the black hole and soliton solutions.  We
first obtain results
for values of the $\sigma$ parameter, defined in (\ref{sigmadef}),
lying in the range $0<\sigma\le 1$.  In section 5 we extend these calculations
to values of $\sigma$ outside the $0<\sigma\le 1$ range, showing how some
new features can now arise.  For example, we find that at certain values of
$\sigma$, the confluence of generically-distinct inverse powers of $r$
can lead to the occurrence of logarithmic radial coordinate dependence in
the solutions, which can then require new kinds of counterterm in order to
cancel divergences.  In section 6, we carry out some numerical studies,
in order to see how the asymptotic forms of the solutions we studied so
far match onto the short-distance forms that arise either near the horizon,
in the case of black holes, or near the origin, in the case of solitons.
The paper ends with conclusions in section 7.  Some details of the exact
static and spherically-symmetric solutions of the Proca equation in
AdS are given in an appendix.

\section{Einstein-Proca AdS Black Holes}\label{setupsec}

   We shall study black hole and soliton solutions in the $n$-dimensional
Einstein-Proca
theory of a massive vector field coupled to gravity, together with a
cosmological constant.  The Lagrangian, viewed as an $n$-form in
$n$ dimensions, is given by
\be
{\cal L} = R\, {*\oneone} + (n-1)(n-2) \ell^{-2}\,  {*\oneone}
    - 2{*F}\wedge F - 2\tilde m^2\,  {*A}\wedge A\,.\label{bulkLag}
\ee
This gives rise to the equations of motion
\bea
E_{\mu\nu}&\equiv& R_{\mu\nu} -
   2 \big(F_{\mu\nu}^2 - \fft{1}{2(n-2)}\, F^2\, g_{\mu\nu}\big)
- 2 \tilde m^2\,  A_\mu A_\nu + (n-1) \ell^{-2}\, g_{\mu\nu}=0\,,
\label{Einsteineom}\\
d{*F} &=& (-1)^n\, \tilde m^2\, {*A}\,.\label{Procaeom}
\eea

  We shall consider spherically-symmetric solutions of the Einstein-Proca
system, described by the ansatz
\be
ds^2 = - h(r) dt^2 + \fft{dr^2}{\th(r)} + r^2 d\Omega_{n-2}^2\,,\qquad
A= \psi(r) dt\,,\label{solans}
\ee
where $d\Omega_{n-2}^2$ is the metric of the unit $(n-2)$-sphere.
The Ricci tensor of the metric in (\ref{solans}) is given by
\bea
R_{tt} &=& h\th\Big(\fft{h''}{2h} - \fft{h'^2}{4h^2} +
\fft{h'\th'}{4h\th} + \fft{(n-2)h'}{2rh}\Big)\,,\nn\\
R_{rr} &=& -\fft{h''}{2h} + \fft{h'^2}{4h^2} - \fft{h'\th'}{4h\th}
- \fft{(n-2)\th'}{2r\th}\,,\nn\\
R_{ij} &=& \Big( (n-3) - \fft{r(h\th)'}{2h} - (n-3) \th\Big) \tilde g_{ij}\,,
\eea
where $\tilde g_{ij}$ is the metric of the unit $(n-2)$-sphere.  The
equations of motion following from (\ref{Einsteineom}) and (\ref{Procaeom})
can then be written as
\bea
&&\psi^2=\fft{n-2}{4\tilde m^2 r}(\th h' - h \th')\,,
\qquad (E_t{}^t - E_r^r=0)
\,,\nn\\
&&-\fft{r(h\th)'}{2h} - \fft{2r^2\th \, {\psi'}^2}{(n-2)h} +
(n-3)(1-\th) + (n-1)\ell^{-2}\,  r^2=0\,,\qquad (E_i{}^j=0)\,,\nn\\
&&\fft{\sqrt{h\th}}{r^{n-2}} \Big(r^{n-2} \sqrt{\fft{\th}{h}} \psi'\Big)'
=\tilde m^2 \, \psi\,,\qquad ({\rm Proca \ eom})\,.\label{einstproceom}
\eea
The remaining equation, which may be taken to be $E_t{}^t=0$, is
in consequence automatically satisfied.

   We shall be interested in studying two kinds of solutions of these
equations, namely black holes and solitons.  In the black hole
solutions the functions $h(r)$ and $\th(r)$ will both vanish on the
horizon.  The Proca equation
implies that the potential $\psi$ will vanish on the horizon also.
In the solitonic solutions the radial coordinate runs all the way down
to $r=0$, which behaves as the origin of spherical polar coordinates.
The functions $h$, $\th$ and $\psi$ all approach constants at $r=0$.

   It is not hard to see, by constructing power-series solutions of the
equations near infinity, and also in the vicinity of a putative horizon
at $r=r_0$, where one
assumes that $h(r_0)=0$ and $\th(r_0)=0$, that black holes could be
expected to arise.  Of course these series expansions do not settle
the question of precisely how the interior and exterior solutions join
together.  This can be studied by means of a numerical integration of
the equations, and we shall discuss this in greater detail in
section \ref{numsec}.  For now, we just remark that the numerical analysis
indeed confirms the existence of black-hole solutions.

    Our immediate interest is in studying the asymptotic behaviour of the
solutions, with a view to seeing how to calculate the mass of the black holes.
This will also be relevant for studying the thermodynamics of the solutions.
As we discussed in the introduction, a rather straightforward case arises if
the mass $\tilde m$ of the Proc field is chosen so that the index $\sigma$,
defined by (\ref{sigmadef}), is equal to 1.  This means that the
inverse powers of $r$ in the asymptotic expansions of the Proca and metric
functions take the simple form given in (\ref{sig1exp}).  It is achieved
by taking the mass of the Proca field to be given by
\be
\tilde m^2 = -\ft14 (n-2)(n-4)\ell^{-2}\,.\label{Procamass}
\ee
One can now systematically solve for the coefficients in the expansions
(\ref{sig1exp}) in terms of the free parameters $q_1$, $q_2$ and $m_2$.
We find that the leading coefficients are given by
\be
n_1=\ft12 m_1=\fft{n-4}{n-2}\, q_1^2\,,\qquad
  n_2=m_2 - \fft{2(n-4)}{n-1}\,  q_1 q_2\,.\label{coeffs1}
\ee
At the next two orders, we find
\bea
q_3 &=& \ft18 (n-2)(n-4)\ell^2\, q_1\,,\qquad
n_3= q_2^2 +\ft14 (n-2)(n-4)\ell^2\, q_1^2\,,\nn\\
m_3&=&\fft{2(n-2)\, q_2^2}{n} +
   \fft{(n-4)(n^2-6n+10)\, \ell^2\, q_1^2}{2(n-2)}\,,\nn\\
q_4&=& \ft1{24} (n-2)(n-4)\ell^2\, q_2\,,\qquad
n_4= \ft13 (n-1)(n-4)\ell^2\, q_1 q_2\,,\nn\\
m_4 &=& \fft{(n-4)(2n^3 -8n^2 + 7n + 11)\ell^2\, q_1 q_2}{3(n^2-1)}
\,.
\eea
It is straightforward to continue the process of solving for the
further coefficients to any desired order.  In fact, for the purposes of
computing the mass and deriving the first law of thermodynamics, it turns
out to be unnecessary to go beyond the orders give in (\ref{coeffs1}).

   It should be noted that the specific choice (\ref{Procamass}) gives a rather
natural higher-dimensional generalisation of the ordinary massless
Einstein-Maxwell system in four dimensions.  Setting $n=4$ we have $\tilde m=0$
and $\alpha=0$, and the Maxwell potential $A=\psi dt$ is simply given
by
\be
\psi= q_1 + \fft{q_2}{r}\,.\label{n4case}
\ee
Thus in this case $q_2$ is the ordinary electric charge of the Reissner-Nordstr\"om AdS black hole in four dimensions, and $q_1$ represents
an arbitrary constant shift in the gauge potential.

   In the more general case where the Proca mass is not fixed to the
special value (\ref{Procamass}) that implies $\sigma=1$, but is instead
allowed to lie anywhere in the range that corresponds to $\sigma\le 1$
(see (\ref{sigmadef})), the leading orders in the asymptotic expansions
take the form given in (\ref{asymptotics}).  Specifically, writing
\bea
\psi&=& \fft{q_1}{r^{(n-3-\sigma)/2}} +
    \fft{q_2}{r^{(n-3+\sigma)/2}}
   +\cdots\,,\nn\\
h &=& r^2\, \ell^{-2} + 1 +
  \fft{m_1}{r^{n-3-\sigma}} + \fft{m_2}{r^{n-3}}  \cdots\,,\nn\\
f &=& r^2\, \ell^{-2} + 1 +  \fft{n_1}{r^{n-3-\sigma}}  +
 \fft{n_2}{r^{n-3}} + \cdots\,,\label{sigle1}
\eea
then substituting into the equations of motion we find
\bea
n_1&=& \fft{n-3-\sigma}{n-2}\, q_1^2\,,\qquad
m_1 = \fft{2(n-3-\sigma)}{n-1-\sigma}\, q_1^2\,,\nn\\
n_2 &=& m_2 - \fft{2(n-3-\sigma)(n-3+\sigma)}{(n-1)(n-2)}\, q_1 q_2\,.
\label{gensigres}
\eea
One can continue solving for higher coefficients to any desired order
in $1/r$.  Unlike the $\sigma=1$ case we discussed previously, this will
in general be a somewhat less neatly ordered process, because of the
intermingling of powers of $1/r$ from different sources.  However, it
turns out for our present purposes, of calculating the mass and
deriving the first law of thermodynamics, that the coefficients given
in (\ref{gensigres}) are sufficient.  Note that they reduce to those
given in (\ref{coeffs1}) in the case that $\sigma=1$.

   The free parameter $m_2$ can be thought of as the ``mass parameter'' of
the black hole or soliton; it is associated with the $r^{-(n-3)}$ fall-off
in the
metric coefficients $h(r)$ and $\th(r)$.  However, it is not itself directly
proportional to the physical mass of the object.  In fact, it can be seen
from the expansions (\ref{sigle1}) that there is the potentially troubling
feature that, unlike in a normal asymptotically-AdS black hole, here
the metric functions have terms with a {\it slower} asymptotic fall off
than the ``mass terms,''  namely the $m_1/r^{n-3-\sigma}$ and
$n_1/r^{n-3-\sigma}$ terms.
Naively, these might be expected to give rise to an infinite result for the
physical mass.

   A simple and usually reliable way to calculate the mass of an
asymptotically-AdS black hole is by means of the AMD procedure devised
by Ashtekar, Magnon and Das \cite{ashmag,ashdas}.  This involves integrating
a certain ``electric'' component of the Weyl tensor over the sphere
at infinity in an appropriate conformal rescaling of the metric.  If we
naively apply the AMD procedure to the solutions described above, we
indeed obtain an infinite result for the physical mass.  Rather than pursuing
this further here, we shall instead use a different approach to calculating
the physical mass, using the technique of the holographic stress tensor.
This forms the subject of the next section.

\section{Holographic Energy}

   One way to calculate the mass of an asymptotically-AdS metric is by
constructing the renormalised holographic stress tensor, via the
AdS/CFT correspondence \cite{balakrau,myers,empjohmye,dehsolske}.  
Thus one adds to the bulk
Lagrangian given in
(\ref{bulkLag}) the standard Gibbons-Hawking surface term and the
necessary holographic counterterms.  At this stage it is more
convenient to write the extra Lagrangian terms as scalar densities rather
than as $n$-forms.  Thus in the gravitational sector we have
\cite{empjohmye,dehsolske,kralarsie}
\bea
{\cal L}_{\rm bulk} &=& \fft1{16\pi G}\, 
   \sqrt{-g}\Big[R - F^{\mu\nu} F_{\mu\nu} -
  2 \tilde m^2\, A^\mu A_\mu - (n-1) \ell^{-2}\Big]\,,\label{Lbulk}\\
{\cal L}_{\rm surf} &=& -\fft1{8\pi G}\, \sqrt{-h}\,K\,,\label{Lgh}\\
{\cal L}_{\rm ct} &=& \fft1{16\pi G}\, \sqrt{-h} \,\Big[
-\fft{2(n-2)}{\ell} + \fft{\ell}{(n-3)}\, {\cal R}
+b_2\, \ell^3\,
    \big(\cR_{\mu\nu}\, \cR^{\mu\nu} - \fft{(n-1)}{4(n-2)}\, \cR^2\big)
\nn\\
&& - b_3\, \ell^5\, \Big(\fft{(3n-1)}{4(n-2)} \,
   \cR \cR_{\mu\nu} \cR^{\mu\nu} - \fft{(n^2-1)}{16(n-2)^2}\, \cR^3
  -2 \cR_{\mu\nu\rho\sigma}\, \cR^{\mu\rho}\, \cR^{\nu\sigma} \nn\\
&&+
   \fft{(n-3)}{2(n-2)}\, \cR^{\mu\nu}\, \nabla_\mu\nabla_\nu \cR -
  \cR^{\mu\nu}\square \cR_{\mu\nu} + \fft1{2(n-2)}\, \cR\square\cR\Big)
  +\cdots \Big]\,,\label{Lct}
\eea
where $K=h^{\mu\nu} K_{\mu\nu}$ is the trace of the second fundamental form
$K_{\mu\nu}= -\nabla_{(\mu} n_{\nu)}$, $\cR_{\mu\nu\rho\sigma}$ and
its contractions denote curvatures
in the boundary metric $h_{\mu\nu}= g_{\mu\nu}- n_\mu n_\nu$, and
\be
b_2= \fft{1}{(n-5)(n-3)^2}\,,\qquad b_3= \fft{2}{(n-7)(n-5)(n-3)^3}\,.
\label{b2b3}
\ee
The ellipses in (\ref{Lct}) denote terms of higher order in curvature or
derivatives, which are only needed in dimensions $n>9$. The expressions
in (\ref{Lct}) should only be included when they yield
divergent counterterms.  This means that the terms with coefficient
$c_2$ should be included only in dimensions $n>5$, and those with
coefficient $c_3$ should be included only in dimensions $n>7$.

   We should also include counterterms for the Proca field.  There
is, furthermore, an option also to add a boundary term for the Proca
field, analogous to the Gibbons-Hawking term ${\cal L}_{\rm surf}$ for
the gravitational field.  Thus we can take
\be
{\cal L}^A_{\rm surf} = -\fft{\gamma}{8\pi G}\, \sqrt{-h}\,
n^\mu F_{\mu\nu}\, A^\nu \,.\label{Procasurf}
\ee
In the case of Dirichlet boundary conditions, where the value of the potential
is fixed on the boundary, the coefficient $\gamma$ would be taken to be zero.
For the counterterms, it turns out that for most of the solutions we shall be
interested in we may simply take
\be
{\cal L}^A_{\rm ct}= \fft{e_1}{16\pi G\ell}\, \sqrt{-h} \,
       A^\gamma A_\gamma\,.
\label{Procact}
\ee
We have left the constants $\gamma$ and $e_1$ arbitrary for now.  It will
turn out that one (dimension-dependent)
linear combination of $\gamma$ and $e_1$ will be determined by the
requirement of removing a divergence in the expression for the energy.
The remaining combination then represents an ambiguity in the definition
of the energy, corresponding to the freedom to perform a Legendre
transformation to a different energy variable.

   The variation of the surface and counterterms with respect to the
boundary metric $h_{\mu\nu}$ gives the energy-momentum tensor $T_{\mu\nu}$
of the
dual theory, with 
$T_{\alpha\beta}=(2/\sqrt{-h})\, \delta I/\delta h^{\alpha\beta}$.  
In our case, since we just wish to compute the energy, and
since our metrics are spherically-symmetric and static, many of the terms
that come from the variations of the counterterms turn out to vanish.
In particular, when we compute $T_{00}$ the only surviving contributions
from the variations of the quadratic and cubic curvature terms
will be those coming from the variation of $\sqrt{-h}$.  This greatly
simplifies the calculations.  The upshot is that we may write
\bea
T_{\alpha\beta} &=& \fft1{8\pi G}\, \Big[ K_{\alpha\beta} - K h_{\alpha\beta} -
   (n-2) \ell^{-1}\, h_{\alpha\beta} + \fft{\ell}{n-3}\,
(\cR_{\alpha\beta}-\ft12 \cR h_{\alpha\beta})\cr
\!\!\!&&\!\!\!-
\ft12 b_2\,\ell^3\,
 \big(\cR_{\mu\nu}\, \cR^{\mu\nu} - \fft{(n-1)}{4(n-2)}\, \cR^2\big)
\, h_{\alpha\beta}
+ \ft12 b_3\,  \ell^5 \,
  \Big(\fft{(3n-1)}{4(n-2)} \,
   \cR \cR_{\mu\nu} \cR^{\mu\nu} - \fft{(n^2-1)}{16(n-2)^2}\, \cR^3\cr
\!\!\!&&\!\!\!
  -2 \cR_{\mu\nu\rho\sigma}\, \cR^{\mu\rho}\, \cR^{\nu\sigma}
+
   \fft{(n-3)}{2(n-2)}\, \cR^{\mu\nu}\, \nabla_\mu\nabla_\nu \cR -
  \cR^{\mu\nu}\square \cR_{\mu\nu} + \fft1{2(n-2)}\, \cR\square\cR\Big)
h_{\alpha\beta}\cr
\!\!\!&&\!\!\! +
  (\gamma n^\nu F_{\mu\nu} A^\nu - \ft12 e_1 \ell^{-1}\, A^\gamma A_\gamma)
h_{\alpha\beta}
  -2 \gamma\, n^\mu F_{\mu (\alpha}\, A_{\beta)} +
      e_1 \, \ell^{-1}\, A_\alpha A_\beta
+ \cdots \Big]\,,\label{enmom}
\eea
where the ellipses denote terms that will not contribute to $T_{00}$ for
our solutions, and terms that are needed in dimensions $n>9$.   The holographic
mass is obtained by integrating $T_{00}$ over the volume of the $(n-2)$ sphere
that forms the spatial boundary of the boundary metric.
   The boundary metric for our solutions, and the normal vector to
the boundary, are given simply by
\be
h_{\mu\nu}\, dx^\mu dx^\nu = -h\, dt^2 + r^2\, d\Omega_{n-2}^2\,,\qquad
n^\mu\del_\mu = \sqrt{\th}\,\fft{\del}{\del r}\,.
\ee

We are now ready to insert the asymptotic expansions that we discussed
earlier.  We can use either (\ref{sig1exp}), in the special case $\sigma=1$,
or more generally (\ref{sigle1}) for all the cases with $\sigma\le 1$.
It turns out that the terms displayed in (\ref{sigle1}) are sufficient
for the purpose, with the coefficients $m_1$, $n_1$ and $n_2$ given by
(\ref{gensigres}).

Substituting into (\ref{enmom}), we find that the counterterms subtract out
all divergences, provided that we impose the relation
\be
e_1= (n-3-\sigma)(1-\gamma)\label{erels}
\ee
on the two coefficients $\gamma$ and $e_1$ associated with the counterterms for
the Proca field.  We then find that
the masses of the black holes in dimensions $5\le n\le 9$
are given by\footnote{Our
convention for the definition of mass in dimension $n$ is such that an ordinary
AdS-Schwarzschild black hole, whose metric is given by (\ref{solans}) with
$h=\th= r^2\, \ell^{-2} + 1 - 2m r^{3-n}$, has mass
$M= (n-2) m \omega_{n-2}/(8\pi)$, where
\be
\omega_{n-2}= \fft{2 \pi^{(n-1)/2}}{\Gamma[(n-1)/2]}  \label{Svol}
\ee
is the volume of the unity $(n-2)$-sphere.}
\bea
n=5:&&\qquad M= \fft{3\pi}{8}\, \Big[-m_2 + \Big(\ft23 \sigma \gamma
            + \ft16 (2-\sigma)(6+\sigma)  \Big) q_1 q_2 +
          \ft14\ell^2 \Big]
\,,\cr
n=6:&& \qquad M=\fft{2\pi}{3}\, \Big[-m_2 + \Big(\ft12 \sigma \gamma
 + \ft1{10} (3-\sigma)(8+\sigma)\Big) q_1 q_2\Big]
\,,\cr
n=7:&& \qquad M = \fft{5\pi^2}{16}\,
  \Big[-m_2 + \Big(\ft25 \sigma \gamma +
   \ft1{15} (4-\sigma)(10+\sigma) \Big)q_1 q_2 -\ft18 \ell^4 \Big]\,,
\,,\cr
n=8:&& \qquad M =\fft{2\pi^2}{5}\, \Big[- m_2 +
\Big( \ft13\sigma \gamma + \ft1{21} (5-\sigma)(12+\sigma)\Big) q_1 q_2\Big]
\,,  \cr
n=9:&& \qquad M=\fft{7\pi^3}{48}\, \Big[-m_2 +
\Big(\ft27\sigma \gamma + \ft1{28} (6-\sigma)(14+\sigma)\Big) q_1 q_2
    + \ft5{64} \ell^6\Big]
\,.\label{n59}
\eea
Note that the term proportional to $\ell^{n-3}$ in each odd dimension is
the Casimir energy.  It would be natural to omit this if one wants to
view the mass as simply that of a classical black hole.

    Although the general expressions for the counterterms at the quartic
or higher order in curvatures are not readily available, we can in fact
easily calculate the holographic mass for the static spherically-symmetric
solutions in any dimension.  Any invariant constructed from
$p$ powers of the curvature $\cR_{\mu\nu\rho\sigma}$ of the boundary
metric $h_{\mu\nu}$ will necessarily just be a pure dimensionless number
times $r^{-2p}$, and so the contributions $T^{\rm gct}_{00}$ to $T_{00}$
coming from the gravitational counterterms
to all orders can simply be written as
\be
T^{\rm gct}_{00}= \fft1{8\pi G\, \ell}\, h_{00}\,
   \sum_{p=0}^\infty \fft{c_p\, \ell^{2p}}{r^{2p}}
\,.
\ee
The constants $c_p$ are then uniquely determined by the requirement of
removing all divergences in the holographic expression for the mass.
Together with the contributions from the surface term and the counterterms
for the Proca field, the complete expression for $T_{00}$ for our
metrics is given by
\be
T_{00}= \fft1{8\pi G}\, \Big[ -\fft{(n-2) h \sqrt{\th}}{r} -
  \gamma\, \sqrt{\th}\, \psi\, \psi' + \ft12 e_1\, \ell^{-1}\, \psi^2
    + h\, \ell^{-1}\, \sum_{p=0}^\infty \fft{c_p\, \ell^{2p}}{r^{2p}}\Big]\,.
\ee
 From this, we can calculate the mass of the Einstein-Proca black holes in
arbitrary dimensions, finding in $n$ dimensions
\be
M= \fft{(n-2)\, \omega_{n-2}}{16\pi}\, \Big[ -m_2 +
 (b_1 + b_2\, \gamma) q_1 q_2 \Big] + E^{\rm Casimir}_n\,,
\label{holmass}
\ee
where $e_1=(n-4)(1-\gamma)$ and
\be
b_1 = \fft{2(n-3-\sigma)(2n-4+\sigma)}{(n-1)(n-2)} \,,\qquad
b_2 = \fft{2\sigma}{n-2}\,.
\ee
Note that if we write the energy in terms of the metric parameter
$n_2$ rather than $m_2$, we obtain the simpler expression
\be
M= \fft{(n-2)\, \omega_{n-2}}{16\pi}\, \Big[ -n_2 +
  \fft{2(n-3-\sigma) + 2 \sigma\, \gamma}{(n-2)}\, q_1 q_2\Big]\,.
\label{massn2}
\ee
The Casimir energies are zero for even $n$, while for odd $n$ we find
\be
E^{\rm Casimir}_n = \fft{ (-1)^{(n-1)/2}\, \pi^{(n-3)/2}\,\ell^{n-3}\,
      (n-1)(n-2) \, (n-4)!!}{2^{(n+3)/2}\, ([(n-1)/2]!)^2}\,.
\ee

\section{Thermodynamics from the Wald Formalism}\label{waldsec}

   Wald has developed a procedure for deriving the
 first law of thermodynamics by calculating the variation of a Hamiltonian
derived from a conserved  Noether current.  The general procedure was developed in \cite{wald1,wald2}.  Its application in Einstein-Maxwell theory can be found in \cite{gao}. Starting from a Lagrangian
${\cal L}$, its variation under a general variation of the fields
can be written as
\be
\delta {\cal L} = {\rm e.o.m.} + \sqrt{-g}\, \nabla_\mu J^\mu\,,
\ee
where e.o.m.~denotes terms proportional to
the equations of motion for the fields.  For the theory
described  by (\ref{bulkLag}), $J^\mu$ is given by
\be
J^\mu = g^{\mu\rho}g^{\nu\sigma} (
\nabla_{\sigma}\delta g_{\nu\rho} - \nabla_{\rho}\delta g_{\nu\sigma})
- 4F^{\mu
\nu} \delta A_\nu\,.
\ee
From this one can define a 1-form $J_\1=J_\mu dx^\mu$ and its Hodge dual
\bea
&&\Theta_{\sst{(n-1)}}=(-1)^{n+1}{*J_{\1}} =
\Theta^{\rm grav}_{\sst{(n-1)}} + \Theta^{A}_{\sst{(n-1)}}\,,\cr
&&\Theta^A_{\sst{(n-1)}}=-4(-1)^n {*F}\wedge \delta A\,.
\eea

   We now specialise to a variation that is induced by an infinitesimal
diffeomorphism $\delta x^\mu=\xi^\mu$.  One can show that
\be
J_{\sst{(n-1)}}\equiv \Theta_{\sst{(n-1)}} - i_{\xi} {*L_0} = {\rm e.o.m} -
d{*J_\2}\,,
\ee
where $i_\xi$ denotes a contraction of $\xi^\mu$ on the first index of the
$n$-form ${*L_0}$, and
\be
J_\2=-d\xi_\1 -  4 (i_\xi A) F\,,
\ee
where $\xi_\1=\xi_\mu dx^\mu$.
One can thus define an $(n-2)$-form $Q_{\sst{(n-2)}}\equiv {*J_\2}$, such
that $J_{\sst{(n-1)}}=dQ_{\sst{(n-2)}}$.  Note that we use the subscript
notation ``$(p)$'' to denote a $p$-form. To make contact with the first
law of black hole thermodynamics, we take $\xi^\mu$ to be the time-like
Killing vector that is null on the horizon.  Wald shows that the
variation of the Hamiltonian with respect to the integration constants of
a specific solution is given by
\be
\delta {\cal H}=\fft{1}{16\pi}\delta \int_c J_{\sst{(n-1)}} -
\int_c d(i_\xi \Theta_{\sst{(n-1)}}) =\fft{1}{16\pi}\int
_{\Sigma^{(n-2)}} \Big(\delta Q_{\sst{(n-2)}} -
i_\xi \Theta_{\sst{(n-1)}}\Big)\,,\label{deltaH}
\ee
where $c$ denotes a Cauchy surface and $\Sigma^{(n-2)}$ is its boundary,
which has two components, one at infinity and one on the horizon.
In particular
\be
\delta Q^A_{\sst{(n-2)}} - i_\xi \Theta^A_{\sst{(n-1)}}
=-4 i_\xi A {*\delta F} + 4(-1)^n i_\xi {*F}\wedge \delta A\,.
\ee

   For the case of our ansatz (\ref{solans}), we find
\bea
Q^{\rm grav}_{\sst{(n-2)}} &=& h'
\sqrt{\fft{\th}{h}}\, r^{n-2} \Omega_{\sst{(n-2)}}\,,\nn\\
Q^{\rm A}_{\sst{(n-2)}} &=& -4\psi \psi' \sqrt{\fft{\th}{h}}\, r^{n-2}
   \Omega_{\sst{(n-2)}}\,,\nn\\
i_\xi \Theta^{\rm grav}_{\sst{(n-1)}} &=& r^{n-2} \Big(
\delta\big(h'\sqrt{\fft{\th}{h}}\big) + \fft{n-2}{r} \,
\sqrt{\fft{h}{{\th}}} \delta\th \Big)\Omega_{\sst{(n-2)}}\,,\nn\\
i_\xi \Theta^{\rm A}_{\sst{(n-1)}} &=& -4(\delta \psi) \psi'
 \sqrt{\fft{\th}{h}}\, r^{n-2} \Omega_{\sst{(n-2)}}\,.
\eea
In the asymptotic region at large $r$, this gives
\be
\delta Q - i_\xi \Theta = r^{n-2}\,\sqrt{\fft{h}{f}}\,
 \Big( -\fft{n-2}{r} \,\delta \th -
\fft{4f}{h}\,
\psi\delta \psi' -2\psi \psi'(\fft{\delta \th}{h} - \fft{\th\delta h}{h^2})
\Big) \Omega_{\sst{(n-2)}}\,.\label{delta}
\ee
 From the boundary on the horizon, one finds
\be
\fft1{16\pi}\int_{r=r_0}(\delta Q - i_\xi \Theta) =
T \delta S\,.\label{TdSterm}
\ee
(This result is a generalisation of that for Einstein-Maxwell theory
obtained \cite{gao}.  Analagous results were obtained in
\cite{Liu:2013gja} for Einstein gravities coupled
to a conformally massless scalar.)

Substituting the asymptotic expansions (\ref{sigle1}) into (\ref{delta}),
we find from (\ref{deltaH}) that the contribution to $\delta{\cal H}$ from
the boundary at infinity is given by sending $r$ to infinity in the
expression
\bea
\delta{\cal H}_\infty &\longrightarrow& \fft{\omega_{n-2}}{16\pi}\Big\{
r^\sigma\, [-(n-2) \delta n_1 + (n-3-\sigma) \delta q_1^2]\cr
&& -
(n-2) \delta n_1 + 2(n-3-\sigma) q_2\delta q_1 + 2 (n-3+\sigma) q_1 \delta q_2
\Big\}\,.
\eea
The ostensibly divergent $r^\sigma$ term in fact vanishes, by virtue of
the relation between $n_1$ and $q_1$ given in (\ref{gensigres}).

If we now use our expression (\ref{massn2}) for the holographic mass,
the variation $\delta {\cal H}_\infty$
can be rewritten as
\be
\delta {\cal H}_\infty = \delta M + \fft{\sigma\, \omega_{n-2}}{4\pi}\,
   q_1 \delta q_2 - \fft{\sigma \gamma\, \omega_{n-2}}{8\pi}\, \delta (q_1 q_2)\,,
\ee
and so, together with (\ref{TdSterm}), we obtain the first law of
thermodynamics in the form
\be
dM = TdS  -\fft{\sigma\, \omega_{n-2}}{4\pi}\,
   q_1 d q_2   + \fft{\sigma \gamma\,\omega_{n-2}}{8\pi}\,
 d(q_1 q_2)\,.\label{firstlaw}
\ee

    It should be recalled that the parameter $\gamma$ can be chosen freely,
with different choices corresponding to making Legendre transformations which
redefine the mass, or energy, of the black hole by the addition of
some constant multiple of $q_1\, q_2$ (see (\ref{holmass})).  A simple
choice is to take $\gamma=0$, in which case we have
\be
dM = TdS - \fft{\sigma\, \omega_{n-2}}{4\pi}\, \, q_1\, d q_2
\,.\label{firstlaw1}
\ee
As mentioned previously, taking $\gamma=0$ corresponds to the case where
the potential at infinity is held fixed in the variational problem.

  It is worth remarking that if we specialise to $\sigma=1$ in four dimensions,
as we observed earlier, then $\tilde m^2=0$ and 
our Einstein-Proca model reduces to the ordinary
Einstein-Maxwell system, then the $q_1\, dq_2$ term in (\ref{firstlaw1})
reduces simply to the standard $\Phi dQ$ contribution in the first law
for charged black holes.  Namely, $q_2$ is then electric charge $Q$, and
the potential difference $\Phi=\Phi_h-\Phi_\infty$ is equal to $-q_1$,
since in our general calculation the potential $\psi$ vanishes on the horizon
and here, in four dimensions, the potential at infinity is equal to $q_1$
(see eqn (\ref{n4case})).  More generally, choosing $\tilde m=0$ in
any dimension $n$ gives $\sigma=n-3$. For this Einstein-Maxwell system
we have $A= q_1 +  q_2\, r^{3-n}$, and again the first law (\ref{firstlaw1})
becomes the standard one for a charged black hole, derived in the gauge 
where the potential vanishes on the horizon.

   Another choice for the constant $\gamma$ that might be considered natural
is to choose it so that
the $q_1\, dq_2$ and $q_2\, dq_1$ terms in the first law (\ref{firstlaw})
occur in the ratio such that they vanish if one imposes a dimensionless
relation between $q_1$ and $q_2$.  Since these quantities have length
dimensions $[q_1]= L^{(n-3-\sigma)/2}$ and
$[q_2]= L^{(n-3+\sigma)/2}$, such a relation
must take the form
\be
q_1^{n-3+\sigma}= c\, q^{n-3-\sigma}\,,\label{qrel}
\ee
where $c$ is a dimensionless constant.  Thus if we choose
\be
\gamma= \fft{n-3+\sigma}{n-3}\,,
\ee
then the first law (\ref{firstlaw}) becomes
\be
dM = TdS - \fft{\sigma\, \omega_{n-2}}{8(n-3)\pi}\,
  \Big[(n-3-\sigma)\, q_1\, d q_2 - (n-3+\sigma)\, q_2\, d q_1\Big]
\,,\label{firstlaw2}
\ee
reducing simply to $dM=TdS$ if (\ref{qrel}) is imposed.

   Finally in this section, we remark that the Wald type
derivation of the first law that we discussed above can be applied also
to the case of solitonic solutions to the Einstein-Proca system.  In these
solutions there is no inner boundary, and instead $r$ runs outwards from $r=0$
which is simply like an origin in spherical polar coordinates.  The
behaviour of the metric and Proca functions at large $r$ takes the same
general form as in (\ref{sigle1}).  Thus when we apply the procedures
described earlier in this section, we can derive a first law of thermodynamics
that is just like the one for black holes, except that the
$TdS$ term that came from the integral (\ref{TdSterm}) over the boundary
on the horizon.  If we make the simple choice $\gamma$ defining the
energy of the system, the first law (\ref{firstlaw}) for the black hole
case will simply be replaced by
\be
dM = -
\fft{\sigma\, \omega_{n-2}}{4\pi}\, \, q_1\, d q_2 \,.\label{solitonfirstlaw}
\ee

\section{Solutions Outside $0<\sigma\le1$}

    Until now, our discussion of the solutions to the Einstein-Proca
system has concentrated on the cases where the Proca mass is such that the
index $\sigma$, defined in (\ref{sigmadef}), satisfies $\sigma\le1$.
This was done in order to allow a relatively straightforward and uniform
analysis of the asymptotic structure of the solutions.  However, it
should be emphasised that black hole and soliton solutions of the
Einstein-Proca equations exist also if the index $\sigma$ lies in a
wider range.  Looking at the form of the expansions in (\ref{sigle1}),
we see that the effects of the back-reaction of the Proca field on
the metric components sets in at a leading order of $1/r^{n-3-\sigma}$.
Clearly, if this were to be of order $r^2$ or higher, then the
back-reaction would be overwhelming the $r^2\ell^{-2}$ terms in $h$ and $f$
that establish the asymptotically-AdS nature of the solutions. Thus we can
expect that in order to obtain asymptotically-AdS black holes or solitons,
we should have
\be
\sigma < n-1\,,
\ee
which, from (\ref{sigmadef}), implies that the Proca mass must satisfy
\be
\tilde m^2 < m_*^2\equiv \fft{n-2}{\ell^2}\,.\label{uppermass}
\ee
Thus the full Proca mass range where we may expect to find stable black hole
and soliton solutions is
\be
-\fft{(n-3)^2}{\ell^2} < \tilde m^2 < \fft{n-2}{\ell^2}\,.\label{fullrange}
\ee
Below the Breitenlohner-Freedman bound which forms the lower limit, we
expect the solutions to be unstable against time-dependent perturbations,
on account of the tachyonic nature of the Proca field.

   In the next section, where we carry out a numerical study of the
various solutions, we find that indeed the upper bound in
(\ref{fullrange})
represents the upper limit of where we appear to obtain well-behaved
black hole and soliton solutions.  At the lower end, the numerical
integrations appear to be stable not only for the entire
Breitenlohner-Freedman window of negative $\tilde m^2$ in (\ref{fullrange}),
but also for arbitrarily negative $\tilde m^2$ below this, where the
Proca field has become tachyonic.  Presumably if we were to extend our
numerical analysis to include the possibility of time-dependent
behaviour we would find exponentially-growing timelike instabilities
below the limit in (\ref{fullrange}), but these cannot be seen in the numerical
integration of the static equations that we study here.

   We may thus divide the range of possible values for $\tilde m^2$, the square
of the Proca mass, as follows:

\begin{itemize}

\item[(1)] $0<\tilde m^2 <m_*^2\,;\qquad\qquad (n-3)<\sigma < (n-1)$

\item[(2)] $-\ft14 (n-2)(n-4)\ell^{-2} \le \tilde m^2 <0\,;
                 \qquad\qquad 1\le\sigma< (n-3)$

\item[(3)] $m^2_{\rm BF} < \tilde m^2 < -\ft14 (n-2)(n-4)\ell^{-2}\,;
 \qquad\qquad  0 <\sigma <1$

\item[(4)] $\tilde m^2 ={\bf m}^2_{\rm BF}\,;\qquad \qquad \sigma=0$

\item[(5)] $ \tilde m^2 < {\bf m}^2_{\rm BF}\,;\qquad\qquad
\sigma\ \hbox{imaginary}$

\end{itemize}

  When the Proca mass lies in the range (1), the leading term in
the fall-off in the metric functions $h$ and $f$ due to back reaction from
the Proca field occurs at a positive power of $r$, lying between 0 and 2.
In the range (2), the leading powers of $r$ in the metric functions
due to back reaction are negative, and so the rate of approach to AdS is more
conventional, but there can still be a rather complicated sequence of
back-reaction terms at more dominant orders than the mass term $m_2/r^{n-3}$
in the metric functions.  The range (3) corresponds to the cases we have
already discussed in general, where the leading-order terms in the asymptotic
expansions take the form (\ref{asymptotics}).  A special case arises in (4),
where the Proca mass-squared precisely equals the negative-most limit of
the Breitenlohner-Freedman bound.  Finally, in the range (5), the Proca
mass-squared is more negative than the Breitenlohner-Freedman bound, and
there is tachyonic behaviour.

  It is instructive to examine some examples of the asymptotic behaviour
of solutions where the Proca mass lies in the various ranges outside
the cases in (3) that we have already studied.
As a first example, we shall consider

\medskip
\noindent{\underline{Dimension $n=5$ with $\sigma=\ft32$}}:
\smallskip

  Note that in this example, the mass-squared $\tilde m^2$ of the Proca
field is still negative, with $\tilde m^2= -7/(4\ell^2)$, and it lies
within the range (2) described above.
We find that the asymptotic expansions for the Proca and
metric functions take the form
\bea
\phi &=& \fft{q_1}{r^{1/4}} + \fft{q_2}{r^{7/4}} +
  \fft{7 \ell^2\, q_1}{16 r^{9/4}} + \fft{7 \ell^2\, q_1^3}{120 r^{11/4}} +
  \fft{\ell^2\, q_2}{16 r^{15/4}} +
  \fft{7(16\ell^2\,  m_2\, q_1 +22 \ell^2\, q_1^2 \, q_2 - 9\ell^4\, q_1}{
              2560 r^{17/4}} + \cdots\,,\nn\\
h&=& r^2\, \ell^{-2} + 1 +\fft{2 q_1^2}{5 r^{1/2}} +
   \fft{m_2}{r^2} + \fft{49\ell^2\, q_1^2}{60 r^{5/2}} +
   \fft{7 \ell^2\, q_1^4}{50 r^3} +\cdots\,,\nn\\
f&=& r^2\, \ell^{-2} + 1 +\fft{q_1^2}{6 r^{1/2}} +
   \fft{24m_2-7 q_1\, q_2}{24 r^2} + \fft{35\ell^2\, q_1^2}{48 r^{5/2}} +
   \fft{91 \ell^2\, q_1^4}{720 r^3} +\cdots\,.
\eea
This example illustrates how when $\sigma>1$ we can get an intermingling of
fall-off powers, with the $r^{-9/4}$ descendant of the leading
$q_1\, r^{-1/4}$ term in the expansion for $\psi$ appearing
prior to the first back-reaction term in (\ref{sigle1}), which
is at order $r^{-11/4}$.   Nevertheless, we find that the holographic mass
and the Wald formula for the first law continue to give finite results
which agree with the general expressions (\ref{holmass}) and (\ref{firstlaw}).

  As a second example we consider

\medskip
\noindent{\underline{Dimension $n=5$ with $\sigma=\ft52$}}:
\smallskip

    In this case the Proca mass-squared is positive,
with $\tilde m^2 = + 9/(16\ell^2)$.  It is still less than the upper limit
$\tilde m^2 = m_*^2 = 3/\ell^2$ in five dimensions that we described above,
and so it lies within the range (1).
The asymptotic forms of the Proca and metric functions are
\bea
\phi&=& q_1\, r^{1/4} - \fft{\ell^2\, q_1^3}{8 r^{5/4}} +
\fft{9 \ell^2\, q_1}{16 r^{7/4}} + \fft{q_2}{r^{9/4}} +
  \fft{21 \ell^4\, q_1^5}{128 r^{11/4}} +\cdots\,,\nn\\
h&=& r^2\ell^{-2} -\ft23 q_1^2\, r^{1/2} + 1 -\fft{\ell^2\, q_1^4}{6r} +
  \fft{\ell^2\, q_1^2}{4 r^{3/2}} + \fft{m_2}{r^2} +
   \fft{55\ell^4\, q_1^6}{144 r^{5/2}} +\cdots\,,\nn\\
f&=& r^2\ell^{-2} -\ft16 q_1^2\, r^{1/2} + 1 -\fft{11\ell^2\, q_1^4}{48r} +
  \fft{9\ell^2\, q_1^2}{16 r^{3/2}} + \fft{8m_2 + 3 q_1\, q_2}{8r^2} +
   \fft{229\ell^4\, q_1^6}{576 r^{5/2}} +\cdots\,.
\eea
In this case there is even more intermingling of the orders in the expansions,
with the first back-reaction term in $\psi$, at order $r^{-5/4}$, preceding
a descendant of the leading order $ q_1\, r^{1/4}$ term and also
preceding the second independent solution that begins with $q_2\, r^{-9/4}$.
In the metric functions the first back-reaction term is at order $r^{1/2}$,
which is prior even to the usual constant term of the pure AdS metric
functions $\bar h=\bar f=r^2\ell^{-2} +1$.

   A new feature that arises in this example is that we must now
add further counterterms in the calculation of the holographic mass, in
order to obtain a finite result.  Specifically, we now need to include
terms
\be
{\cal L}_{\rm ct}^{A,\, {\rm extra}} = \fft{e_2\, \ell}{8\pi G}\, \sqrt{-h}\,
    (A^\mu A_\mu)^2  + \fft{e_3\, \ell}{8\pi G}\,\sqrt{-h}\,
   {\cal R}\, A^\mu A_\mu\,.\label{ctextra}
\ee
The divergences in the holographic mass are then removed if we take
\be
e_1= \ft12 (\gamma-1)\,,\qquad e_2= \ft1{18} (3-2 \gamma)\,,\qquad
   e_3 = \ft13 (1-\gamma)\,.
\ee
The resulting expression for the holographic mass then agrees with the
general formula (\ref{holmass}), after specialising to $n=5$ and
$\sigma=\ft52$.  Using the Wald procedure described in section \ref{waldsec},
we obtain the specialisation of (\ref{firstlaw}) to $n=5$ and
$\sigma=\ft52$.

 For another example, we consider

\medskip
\noindent{\underline{Dimension $n=6$ with $\sigma=2$}}:
\smallskip

This
lies within the range of category (2) above, and has
$\tilde m^2 = -5/(4\ell^2)$.  Another new feature arises here,
namely that we find also $\log r$ behaviour in the asymptotic
expansions of the metric and Proca functions.  To the first few orders at
large $r$, we find
\bea
h&=& \ell^{-2} r^2 + 1 + \fft{2q_1^2}{3r} -
 \fft{5\ell^2 q_1^2\log r}{4r^3} + \fft{m_2}{r^3} \cdots\,,\cr
f &=& \ell^{-2} r^2 +1+ \fft{q_1^2}{4r} -
   \fft{15\ell^2q_1^2 \log r}{16r^3} + \fft{n_2}{r^3} + \cdots\,,\cr
\psi &=& r^{-\fft12} \Big(q_1 -\fft{5\ell^2 q_1\log r}{8 r^2}
    + \fft{q_2}{r^2} + \cdots\Big)\,,\cr
&& m_2=n_2 + \ft12 q_1 q_2 - \ft7{48} \ell^2 q_1^2\,.
\eea
The Wald formula still gives a convergent result, with
\be
\fft{16\pi}{\omega_4} \delta {\cal H}_{\rm \infty} = -4 \delta n_2
+2q_2 \delta q_1 + 10 q_1 \delta q_2 + \ft52\ell^2 q_1 \delta q_1\,.
\label{waldn62}
\ee
Now, however, using just the counterterms we have discussed so far we find that
there is still an order  $\log r$ divergence remaining in the
holographic mass.  It can be
removed if we make a specific choice for the coefficient $\gamma$ in
(\ref{Procasurf}), namely $\gamma=1$.  Thus in this case we are effectively
obliged to view the surface term (\ref{Procasurf}) as being instead
a counterterm, which serves the purpose of removing the logarithmic
divergence.
We then find that all the divergences in the holographic mass are removed 
if we take
\be
e_1= 0\,,\qquad \gamma=1\,,
\ee
leading to  
\be
M= \fft{2\pi}{3}\, \Big[ -n_2 + \ft32\, q_1 q_2 +
 \ft1{16}\, (48 e_3 + 5)\, \ell^2\, q_1^2\Big]\,.\label{Mn2}
\ee
Note that although the $\sqrt{-h}\, {\cal R}\, A^\mu A_\mu$ counterterm in
(\ref{ctextra}), with its coefficient $e_3$, is not required for the
purpose of subtracting out any divergence, it is now making
a contribution to the holographic mass, representing an ambiguity
in its definition.  

   If we use (\ref{Mn2}) to eliminate $n_2$ from (\ref{waldn62}), this
leads to the first law
\be
dM=TdS -\fft{2\pi}{3}\, \big[ 3  q_1\, dq_2 +  q_2\, dq_1
   + 6 e_3\, q_1 dq_1\big]\,.
\ee
A natural choice would be to take the free parameter $e_3$ to vanish.

  For the next example, we consider case (4) above, where the Proca mass
satisfies exactly the Breitenlohner-Freedman bound:

\medskip
\noindent{\underline{Dimension $n$ with $\sigma=0$}}:
\smallskip

   In $n$ dimensions, we therefore have $\tilde m^2 = -(n-3)^2/(4\ell^2)$.
The large-distance expansions take the form
\bea
h &=&\ell^{-2} r^2 + 1 + \fft{m_1 (\log r)^2+ \tilde m_1 \log r +
   m_2}{r^{n-3}} + \cdots\,,\cr
f &=& \ell^{-2} r^2 +1+ \fft{n_1 (\log r)^2 +
   \tilde n_1 \log r + n_2}{r^{n-3}} + \cdots\,,\cr
\psi &=& \fft{q_1 \log r + q_2}{r^{(n-3)/2}}+\cdots\,,\qquad
m_1 =2n_1= \fft{2(n-3) q_1^2}{n-1}\,,\cr
\tilde m_1 &=&\fft{4q_1((n-1)(n-3) q_2-2q_1)}{(n-1)^2}\,,\qquad
\tilde n_1 = \fft{2q_1((n-3) q_2-q_1)}{n-2}\,,\cr
m_2 &=&  n_2 +
\fft{(n-3)^2(2q_1^2 + 2(n-1) q_1 q_2 + (n-1)^2 q_2^2)}{(n-1)^3(n-2)}\,.
\label{expansionssig0}
\eea
We see that here also, there is logarithmic
dependence on the $r$ coordinate in the asymptotic expansions.
Nevertheless, the Wald formula turns out to be convergent, and we find
\be
\fft{16\pi}{\omega_{n-2}} \delta {\cal H}_{\rm \infty} =
-(n-2) \delta n_2 - 4 q_2 \delta q_1 + 2(n-3) q_2 \delta q_2\,.
\label{delHsig0}
\ee

   Logarithmic divergences proportional to $\log r$ and
$(\log r)^2$ arise in these $\sigma=0$ examples, and as in 
the earlier case of $\sigma=2$ in
$n=6$ dimensions, it is necessary to make a specific choice for 
the coefficient $\gamma$ in the surface term (\ref{Procasurf}) in
order to remove them, namely by setting
\be
\gamma = \fft{4}{n-2}\,.\label{e1choice}
\ee
Again it turns out that the original Proca counterterm (\ref{Procact}) is
not required in this case, and so we take $e_1=0$ here.  The holographic mass
is then given by
\be
M= \fft{(n-2)\, \omega_{n-2}}{16\pi}\, \Big[-n_2 +\fft1{n-2}\,
  \big((n-3) q_2^2 - 2 q_1 q_2\big)\, q_2\Big]\,,
\ee
and hence from (\ref{delHsig0}) we arrive at the first law
\be
dM = T dS -
  \fft{\omega_{n-2}}{8\pi}\, (q_1\, dq_2 - q_2\,  dq_1)\,.
\ee
The reason why $q_1$ and $q_2$ enter in a rather symmetrical way here may
be related to the fact that in this $\sigma=0$ case the dimensions of
the two quantities $q_1$ and $q_2$ are the same, as can be seen from the
expansion for $\phi(r)$ in (\ref{expansionssig0}).  Also, the
different way in which the logarithmic singularities in the holographic
mass are handled in these $\sigma=0$ examples could be related to the fact the
that the original solution for $\phi$ itself, prior to taking back reactions
into account, already has the logarithmic dependence associated with the
coefficient $q_1$.  Again, this arises because $q_1$ and $q_2$ have the
same dimensions when $\sigma=0$.  By contrast, in an example such as
$n=6$, $\sigma=2$ discussed previously, the logarithmic dependences
arose only via the back reactions, as a result of a confluence of powers of
$r$ in these sub-leading terms.

  Finally, we consider the cases (5)  where the Proca mass-squared is
more negative than the Breitenlohner-Freedman bound. We make take

\medskip
\noindent{\underline{Dimension $n$ with $\sigma=\im\, \tilde\sigma$}}:
\smallskip

Here we have
\be
\tilde\sigma =\sqrt{-\ft14 \tilde m^2 \ell^2 -(n-3)^2 }= -\im\, \sigma\,,
\ee
with $\tilde m^2 < {\bf m}_{\rm BF}^2$ and so $\tilde\sigma$ is real.
We find that the asymptotic expansions take the form
\bea
h &=& \ell^{-2} r^2 + 1 +
\fft{m_1 \cos(\tilde \sigma \log r) + \tilde m_1
\sin(\tilde \sigma \log r) + m_2}{r^{n-3}} + \cdots\,,\cr
f &=& \ell^{-2} r^2 + 1 + \fft{n_1 \cos(\tilde \sigma \log r) +
\tilde n_1
\sin(\tilde \sigma \log r) + n_2}{r^{n-3}} + \cdots\,,\cr
\psi &=& \fft{q_1 \cos (\fft12 \tilde \sigma \log r) +
q_2\sin(\fft12 \tilde \sigma \log r)}{r^{(n-3)/2}} + \cdots\,,\cr
m_1 &=& \fft{(q_1^2-q_2^2)\tilde\sigma^2 - 4q_1q_2 \tilde\sigma +
        (n-1)(n-3)(q_1^2-q_2^2)}{\tilde \sigma^2 + (n-1)^2}\,,\cr
\tilde m_1 &=& \fft{2(q_1^2-q_2^2)\tilde\sigma + 2q_1 q_2((n-1)(n-3)
                   + \tilde \sigma^2)}{\tilde \sigma^2 + (n-1)^2}\,,\cr
n_1 &=& \fft{(n-3)(q_1^2-q_2^2)-2q_1q_2\tilde \sigma}{2(n-2)}\,,\cr
\tilde n_1 &=& \fft{(q_1^2-q_2^2)\tilde\sigma + 2 (n-3)q_1 q_2}{2(n-2)}\,,\cr
m_2 &=& n_2 + \fft{(q_1^2+q_2^2)((n-3)^2 +\tilde\sigma^2)}{2(n-1)(n-2)}\,.
\eea
The resulting Wald formula is given by
\be
\fft{16\pi}{\omega_{n-2}} \delta {\cal H}_{\rm \infty} =
-(n-2) \delta n_2 + \tilde\sigma (q_2\delta q_1 - q_1\delta q_2) +
\ft12(n-3) \delta (q_1^2+q_2^2)\,.\label{delHlast}
\ee
Again, as in the previous example, the original Proca counterterm
(\ref{Procact}) is not required to regularize the holographic mass, and
so we may take $e_1=0$.  Terms proportional
to $\sin(\tilde\sigma\, \log r)$ and $\cos(\tilde\sigma\, \log r)$ can be
removed by taking the coefficient $\gamma$ of the Proca surface term
(\ref{Procasurf}) to be again given by (\ref{e1choice}).  This yields the
finite result
\be
M= \fft{(n-2)\, \omega_{n-2}}{16\pi}\,
\Big[ - n_2 + \fft{(n-3)}{2(n-2)}\, (q_1^2 + q_2^2)\Big]
\ee
for the holographic mass, and hence, from (\ref{delHlast}), we arrive at the
first law
\be
dM = TdS -
  \fft{\tilde\sigma\,  \omega_{n-2}}{16\pi} \, (q_2 d q_1 - q_1 dq_2)\,.
\ee

\section{Numerical Results}\label{numsec}

   It does not appear to be possible to solve the Einstein-Proca
equations of motion for
static spherically-symmetric geometries analytically, and so we now resort
to numerical integration in order to gain more insight into the solutions.
Two distinct kinds of regular solutions can arise; firstly black holes, and
secondly what we shall refer to as ``solitons.''

   The black hole solutions
can be found numerically by first assuming that there exists an horizon
at some radius $r=r_0$, at which the metric functions $h(r)$ and $f(r)$
vanish, then performing Taylor expansions of the
metric and Proca field functions $h(r)$, $f(r)$ and $\psi(r)$ around the
point $r_0$, and then using these expansions to set initial conditions
just outside the horizon for a numerical integration out to infinity.
The criterion for obtaining a ``regular'' black hole solution is that the
functions should smoothly and stably approach the asymptotic forms
(\ref{sig1exp}) that we assumed in our discussion in section
\ref{setupsec}.  Since, as we have seen, the general asymptotic solutions,
with all three independent parameters $q_1$, $q_2$ and $m_2$ nonvanishing,
are well-behaved at infinity (provided the Proc mass $\tilde m$
satisfies $\tilde m^2<m_*^2$, where $m_*$ is defined in eqn (\ref{uppermass}),
there is no reason why a solution that is well-behaved on the horizon
will not integrate out smoothly to a well-behaved solution at infinity,
and indeed, that is what we find in the numerical analysis.\footnote{Note
that the situation would be very different in the absence of a cosmological
constant.  The asymptotic form of the Proca solutions is then given by
(\ref{psimink}), and so one of the two solutions diverges exponentially
at infinity, assuming $\tilde m^2>0$.  The analogous evolution
from a well-behaved starting-point on the horizon would then inevitably
pick up the diverging solution at infinity, leading to a singular
behaviour.  The exponential divergence could be avoided if $\tilde m^2$ were
negative, but in a Minkowski background this would always be tachyonic, and
so there would be instabilities because of
exponential run-away behaviour as a function of time.}

   The solitonic solutions have a very different kind of interior behaviour,
in which the functions $h(r)$, $f(r)$ and $\psi(r)$ all approach
constant values at the origin at $r=0$.  To study these numerically we
start by obtaining small-$r$ expansions for the functions, using these to
set initial conditions just outside the origin, and then integrating out
to large $r$.  The asymptotic forms of the metric and Proca functions
will again be of the general form given in (\ref{sig1exp}) in the
case of smooth solitonic solutions.  Again, since the generic asymptotic
solutions with Proca mass satisfying $\tilde m^2 <m_*^2$ are well-behaved,
the smooth solutions near the origin will necessarily evolve to solutions
that are well-behaved at infinity.  Since the solitonic
solutions are somewhat simpler than the black holes, we shall begin first
by investigating the solitons.  For the rest of this section, we shall,
without loss of generality, set the AdS scale size by taking
\be
\ell=1\,.
\ee

\subsection{Solitonic solutions}

   The soliton solutions we are seeking have no boundary at small $r$; rather,
$r=0$ will be like the origin of spherical polar coordinates.  We begin
by making Taylor expansions for the metric and Proca functions, taking the
form
\bea
&&h =\alpha (1 + b_2 r^2 + b_4 r^4 + \cdots)\,,\qquad
\th =1 + c_2 r^2 + c_4 r^4 + \cdots\,,\cr
&&\psi=\sqrt{\alpha} (a_0  +a_2 r^2 + a_4 r^4 + \cdots)\,.\label{solitonexp}
\eea
Substituting into the equations of motion (\ref{einstproceom}), one can
systematically solve for the coefficients $(a_2, a_4,\ldots)$,
$(b_2,b_4,\ldots)$ and $(c_2,c_4,\ldots)$ in terms of the coefficient $a_0$.
Thus we find
\bea
a_2 &=& - \fft{(n-2)(n-4)a_0}{8(n-1)}\,,\qquad
b_2 =\fft{2(n-1)\ell^{-2} - (n-2)(n-4)a_0^2}{2(n-1)}\,,\nn\\
c_2&=&\fft{2(n-1)\ell^{-2} + a_0^2(n-4)}{2(n-1)}\,,
\eea
with progressively more complicated expressions for the higher coefficients
that we shall not present explicitly here.
The coefficient $\alpha$ represents the freedom to rescale
the time coordinate.  In an actual numerical calculation, when we
integrate out
to large $r$, we may, without loss of generality, start out by choosing
$\alpha=1$ , and then, by taking the limit
\be
\lim_{r\rightarrow\infty} \fft{h}{r^2} = \beta\,,
\ee
determine the appropriate scaling factor that allows us the to
redefine our $\alpha$ by setting $\alpha =1/\beta$.
Thus we see that the soliton solution has only the one free parameter,
 $a_0$.

   To illustrate the numerical integration process, let us consider the
example of the soliton in dimension $n=5$.  We determined the coefficients
in the expansions (\ref{solitonexp}) up to the $r^8$ order, and used these
to set initial conditions for integrating the equations (\ref{einstproceom})
out to large $r$.  We found that indeed the solitons, parameterised by
the constant $a_0$,  have a well-behaved and stable asymptotic behaviour,
in which the functions $h$, $\th$ and $\psi$ approach the forms
given in (\ref{sig1exp}) with $n=5$.  In particular, we have
\be
h\rightarrow r^2 + 1 + \fft{m_1}{r} + \fft{m_2}{r^2}+\cdots\,,\qquad
\psi \rightarrow \fft{q_1}{r^{1/2}} + \fft{q_2}{r^{3/2}} +\cdots\,.
\label{n5asymptotic}
\ee
(Recall that we are setting the AdS scale $\ell=1$ in this section.)

   In principle, for a given solitonic solution determined by the choice of
the free parameter $a_0$, we can match the numerically-determined asymptotic
form of the solution to the expansions (\ref{n5asymptotic}), and hence
read off the values of the coefficients $q_1$, $q_2$, $m_1$ and $m_2$.  It is
quite delicate to do this, especially to pick up the coefficient $m_2$ which
occurs at four inverse powers of $r$ down from the leading-order
behaviour of $h(r)$.  A useful guide to the accuracy of the
integration routine is to match the numerical results for $h(r)$ at large
$r$ to an assumed form
\be
h= \gamma_0\, r^2 + \gamma_1 + \fft{m_1}{r} + \fft{m_2}{r^2}\,.
\ee
Ideally, one should find $\gamma_0=1$ and $\gamma_1=1$.  There will in fact, of
course, be errors.  We first rescale the numerically-determined $h(r)$
and $\psi(r)$ by the factors $1/\gamma_0$ and $1/\sqrt{\gamma_0}$
respectively.  A test of a reliable solution is then that to high
accuracy we should find (see (\ref{coeffs1}))
\be
\gamma_1 -1 \sim 0\,,\qquad m_1- \ft23 q_1^2\sim 0\,.
\ee

\begin{figure}[ht]
\ \ \ \ \ \includegraphics[width=7cm]{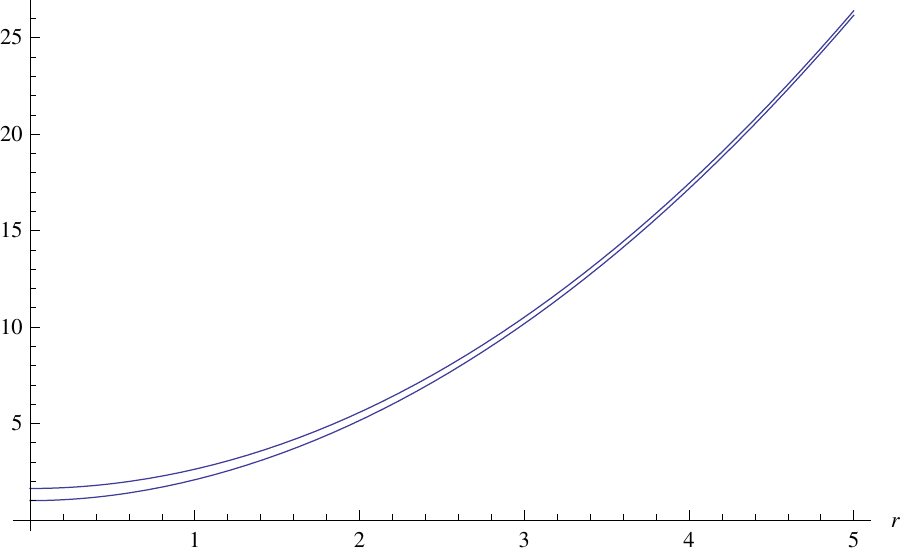}\ \ \ \
\includegraphics[width=7cm]{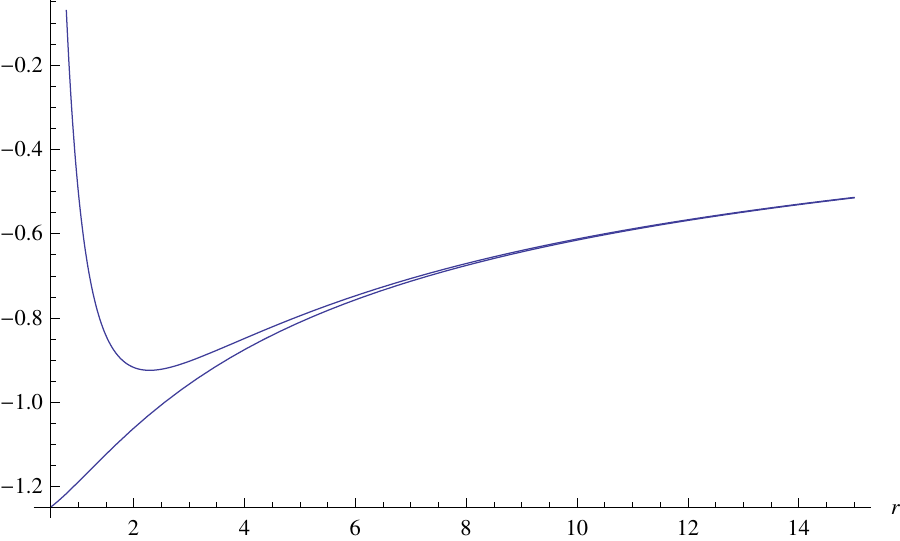}
\caption{\it Smooth soliton in $n=5$ dimensions, with $a_0=-1$ and
$\alpha=1.6285$.  The left-hand plot shows the metric functions
$h(r)$ and $f(r)$.  The upper line is $h(r)$, starting from
$h(0)=\alpha$, and the lower line is $f(r)$, starting from $f(0)=1$.
To leading order, they coalesce at large $r$.  The right-hand plot
shows the potential function $\psi$ (lower line), and compares it
with a best fit of the function
$\tilde\psi = q_1/r^{\fft12} + q_2/r^{\fft32}$ (upper line) that represents the
leading-order terms in the large-$r$ expansion in (\ref{solitonexp}),
achieved by taking
$q_1=-2.0980$ and $q_2=1.6018$.  The numerically-obtained potential
function $\psi$ converges at small $r$ to $\psi(0)=\sqrt{\alpha}$.}
\label{specificsoliton}
\end{figure}

    We now present an explicit example of a smooth soliton solution,
for which we shall make the choice  with $a_0=-1$ in (\ref{solitonexp}).
(We choose a negative value of $a_0$ so that $q_2$ is positive, and in fact
$q_2>0>q_1$.)  For this choice of $a_0$, we find that the free scaling
parameter $\alpha$ should be chosen to be $\alpha=1.6285$ in order to ensure
that $\beta =1.0000$.  The behaviour of the metric functions $(h,f)$ and
the potential function $\psi$ is then displayed in
Fig~\ref{specificsoliton}.  In the left-hand plot, can be seen that
the metric functions are indeed running smoothly from constant values at the
origin, and at large $r$ they approach their expected AdS forms.
By a careful matching of the asymptotic forms of the function $h(r)$ to
the expansion (\ref{solitonexp}), we can read off
a value for the ``mass'' parameter $m_2$, finding  $m_2=-6.2630$.

Repeating this calculation for a range of values for $a_0$, we obtain
reasonably trustworthy numerical expressions for
\be
m_2= m_2(a_0)\,,\qquad q_1=q_1(a_0)\,,\qquad q_2 = q_2(a_0)\,.
\ee
We can then choose to use $q_2$ to parameterise the solutions,
 rather than $a_0$, so that we can write
\be
m_2 = m_2(q_2)\,,\qquad q_1= q_1(q_2)\,.
\ee

\begin{figure}[ht]
\ \ \ \ \ \includegraphics[width=7cm]{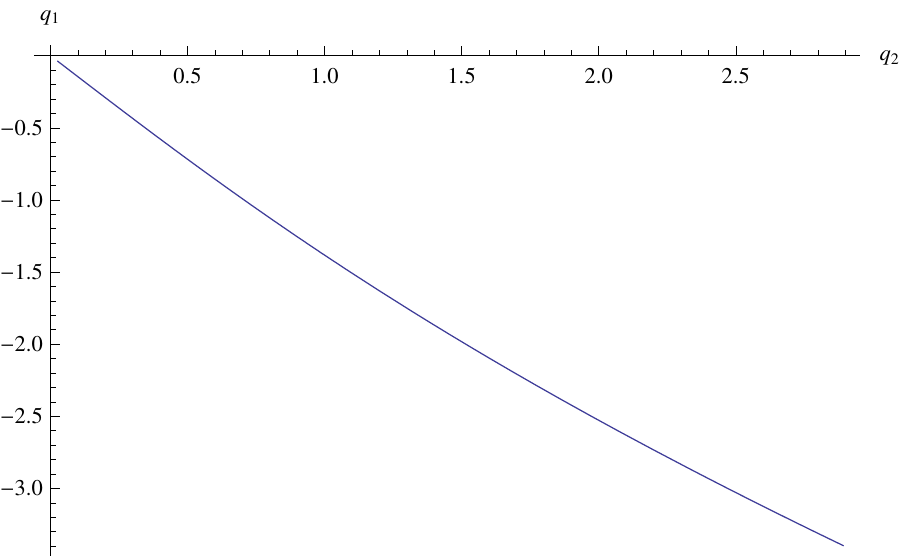}\ \ \ \
\includegraphics[width=7cm]{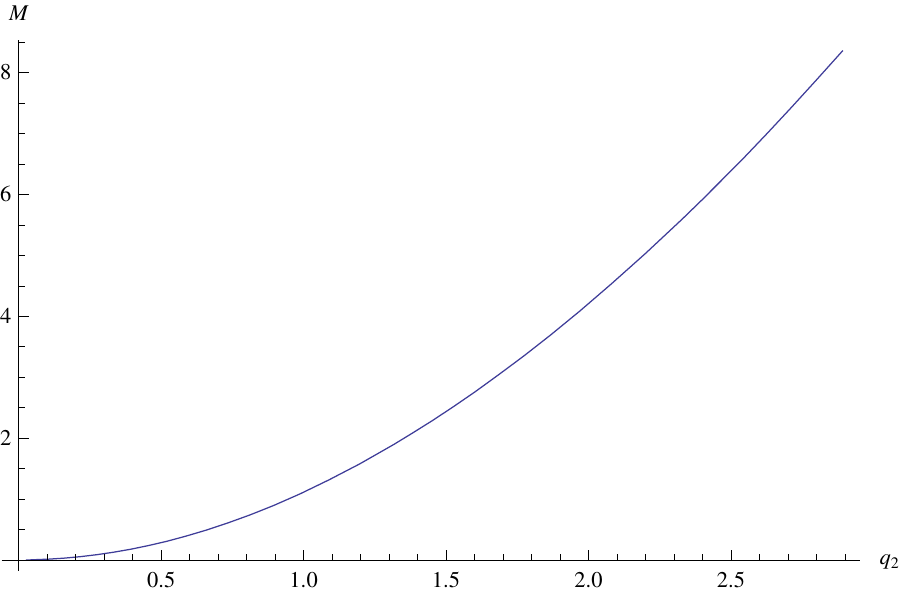}
\caption{\it The left-hand plot shows $q_1$ as a function of $q_2$, whilst
the right-hand plot shows the mass $M$ as a function of $q_2$,
for $0<q_2<3$.  Correspondingly, $a_0$ runs from $a_0=0+$ to $a_0=-1.3$.}
\label{solitonmq1q2}
\end{figure}

   We are now in a position to attempt a numerical verification of the
first law of thermodynamics for the solitonic solutions.  Choosing the
parameter $\gamma$ in section \ref{waldsec} to be zero for simplicity, the
energy of the five-dimensional soliton is given, from (\ref{n59}), by
\be
M= \fft{3\pi}{8}\, \big(-m_2 + \ft76 q_1\, q_2\big)\,.
\ee
Viewing $M$ and $q_1$ as functions of $q_2$, we can now check how
accurately the first law (\ref{solitonfirstlaw}), which in $n=5$ dimensions
reads
\be
dM = -\ft12\pi\, q_1 \, dq_2\label{fls}
\,,
\ee
is satisfied.  In Fig.~\ref{solitonmq1q2}, we display plots of
$q_1(q_2)$ and $M(q_2)$ for a range of $q_2$ values with $0<q_2<3$.

After fitting the data for these curves, we obtain the approximate relations
\be
q_1=-1.486 q_2 + 0.1094 q_2^2\,,\qquad
M=\ft12\pi (0.7449 q_2^2 - 0.03764 q_2^3)\,.\label{q1Mrels}
\ee
Thus we find
\be
-\fft{2}{\pi}\fft{\partial M}{\partial q_2} =
- 1.490 q_2 + 0.1129 q_2^2\,,
\ee
which should, according to (\ref{fls}), be equal to $q_1$.  It is indeed
in reasonable agreement with the approximate form for $q_1$ given in
(\ref{q1Mrels}).

It is worth pointing out that as we increase the negative value of $a_0$,
the mass becomes divergent, with the solution becoming
singular around $a_0=-1.8$.

\subsection{Black hole solutions}

    The static black hole solutions that we are seeking are
characterised by the fact that the Killing vector $\del/\del t$ will become
null on the horizon at $r=r_0$.  Thus the metric function $h(r)$ in
(\ref{solans}) will have a zero at $r=r_0$.  It follows from the equations of
motion (\ref{einstproceom}) that the functions $\th(r)$ and $\psi(r)$ will
vanish at $r=r_0$ also.  We are therefore led to consider near-horizon
series expansions of the form
\bea
&&h=b_1\Big[(r-r_0) + b_2 (r-r_0)^2 + \cdots\Big]\,,\nn\\
&&\th=c_1(r-r_0) + c_2(r-r_0)^2 + \cdots,,\nn\\
&&\psi= \sqrt{b_1}\Big[a_1 (r-r_0) + a_2 (r-r_0)^2+ \cdots\Big]\,.
\label{horexp}
\eea
The constant $b_1$ parameterises the freedom to rescale the time coordinate
by a constant factor. It can be used in order to rescale the solution,
after numerical integration out to large distances, so that the time
coordinate is canonically normalised.

   Substituting the expansions
 (\ref{horexp}) into the equations of motion (\ref{einstproceom}), we
can systematically solve for the coefficients $(a_2,a_3,\ldots)$,
$(b_2,b_3,\ldots)$ and $(c_1,c_2,\ldots)$ in terms of $r_0$ and $a_1$.  In
dimension $n$ the solution for $c_1$ is
\be
c_1=\fft{(n-2)(n-3)+(n-1)r_0^2 }{(2r_0 a_1^2 + n-2) r_0}
\,.\label{c1a1}
\ee
(Recall, again, that we have set the AdS scale $\ell=1$ in this section.)
The expressions for the higher coefficients are all quite complicated
in general dimensions.  Here, as an example,
we just present $(a_2,b_2,c_2)$ in the
special case of $n=5$:
\bea
a_2 &=& -\fft{a_1(4r_0^4 a_1^4 + 12 r_0^3 a_1^2 + 153 r_0^2 + 72)}{
                   48r_0(2r_0^2+1)}\,,\nn\\
b_2 &=& -\fft{2r_0^4 a_1^4 - 93 r_0^3 a_1^2 - 32 r_0 a_1^2+ 24 r_0^2 +36}{
                24 r_0 (2r_0^2 + 1)}\,,\nn\\
c_2 &=& \fft{6r_0^4 a_1^4 +105 r_0^3 a_1^2 + 32 r_0 a_1^2- 24 r_0^2 -36}{
             4 r_0^2 (2r_0^2a_1^2 + 3)}\,.
\eea
In our actual numerical calculations, we have expanded up to and including
the $(r-r_0)^4$ order.  These expansions are then used in order to set
initial conditions just outside the horizon.  We then integrate out to
large $r$.  The criterion for a good black hole solution is that the
metric and Proca functions should approach the asymptotic forms given in
(\ref{sig1exp}).  We find that indeed such solutions arise, and
they are stable as the parameters $r_0$ and $a_1$ are adjusted.

   As we did in the case of the solitonic solutions, here too we can attempt
a numerical confirmation that these black hole solutions obey the first law
of thermodynamics that we derived in section \ref{waldsec}.  Taking the simple
choice $\gamma=0$ again, the first law is given by (\ref{firstlaw1}).
We shall present an example calculation in $n=5$ dimensions, for which the
first law becomes
\be
dM = TdS-\ft12\pi\, q_1 \, dq_2\,.
\ee
The easiest case to consider is when we fix the entropy.
This corresponds to holding $r_0$ fixed, and the first law becomes
\be
dM =  TdS - \ft12 \pi q_1 dq_2\qquad \longrightarrow
\qquad dM = -\ft12 \pi q_1  dq_2\,.
\ee

\begin{figure}[ht]
\ \ \ \ \ \includegraphics[width=7cm]{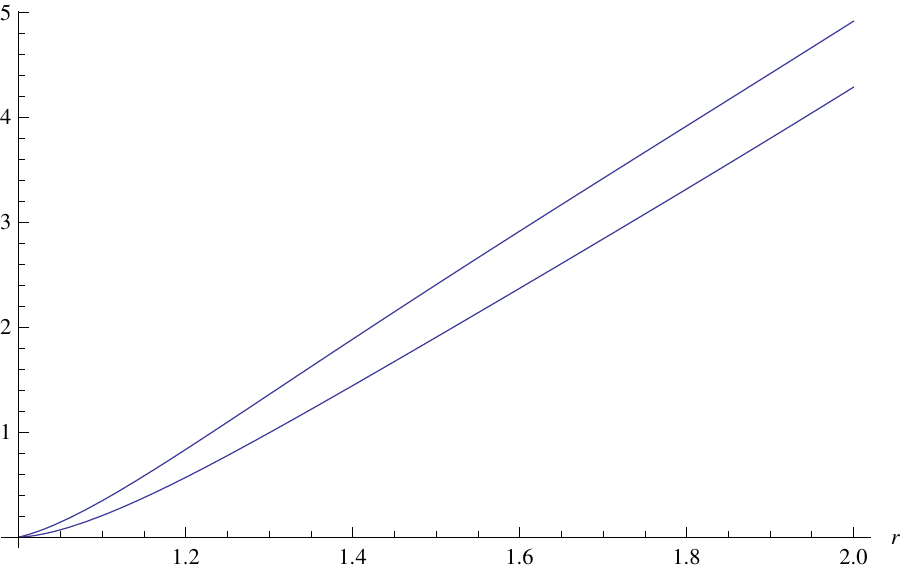}\ \ \ \
\includegraphics[width=7cm]{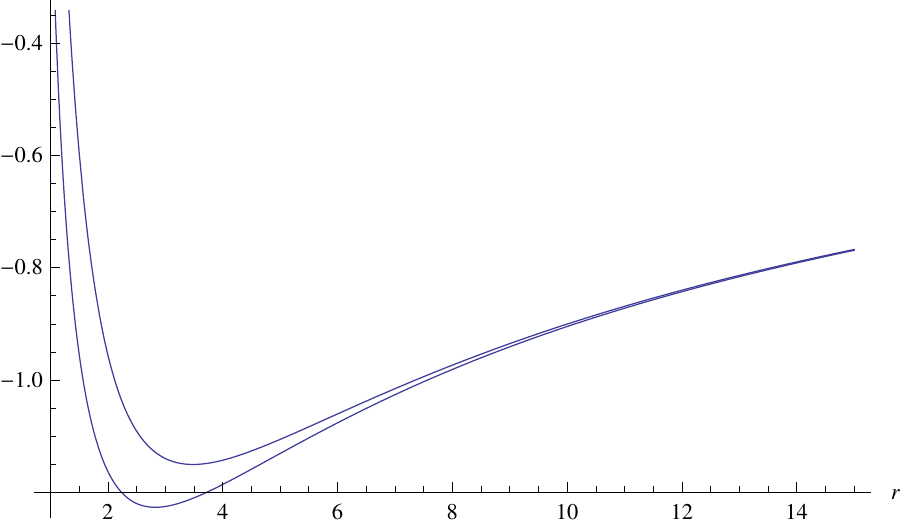}
\caption{\it Black hole in $n=5$ dimensions, with $a_1=-10$ and
$b_1=2.163$.  The left-hand plot shows the metric functions
$h(r)$ and $f(r)$.  The upper line is $h(r)$, starting from
$h(r_0)=0$, and the lower line is $f(r)$, starting from $f(r_0)=0$.
To leading order, they coalesce at large $r$, which we did not present
in this figure.  The right-hand plot
shows the potential function $\psi$ (lower line), and compares it
with a best fit of the function
$\tilde\psi = q_1/r^{\fft12} + q_2/r^{\fft32}$ (upper line) that
represents the
leading-order terms in the large-$r$ expansion in (\ref{solitonexp}),
achieved by taking
$q_1=-3.222$ and $q_2=3.742$.  The actual numerically-obtained potential
function $\psi$ vanishes at small $r=r_0$, although $\tilde \psi(r_0)\ne 0$.}
\label{specificbh}
\end{figure}

   As a concrete example, let us set $r_0=1$. The solution then has
one non-trivial adjustable parameter, $a_1$, remaining.  The parameter
$b_1$ should
be fixed so that we have $\fft{h}{r^2}|_{r\rightarrow \infty} = 1$ at large
$r$.  To see in more detail how the metric function and $\psi$ behave,
let us take as an example $a_1=-10$, implying $c_1=18/203$.
We find that we should then take $b_1=2.163$.  Thus the black hole has
a temperature
\be
T=\fft{b_1 c_1}{4\pi} = 0.01526\,.
\ee
The plots for the metric functions $(h,f)$ and the Proca potential
$\psi$ are given in Fig.~\ref{specificbh}.  From the numerical solution,
we can read off $m_2=-23.93$ and $(q_1,q_2)=(-3.222,3.742)$, and
hence find that the mass is given by $M=11.63$.

\begin{figure}[ht]
\ \ \ \ \includegraphics[width=7cm]{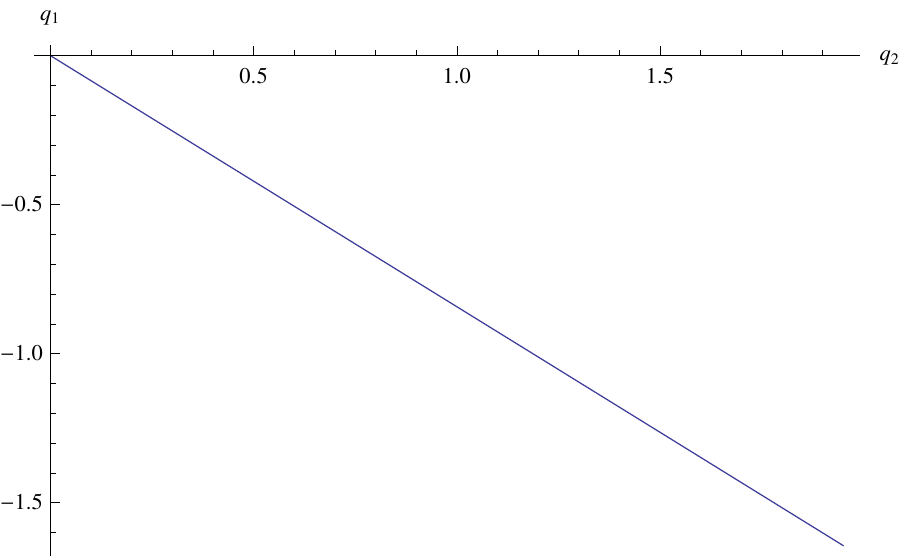}\ \ \ \ \
\includegraphics[width=7cm]{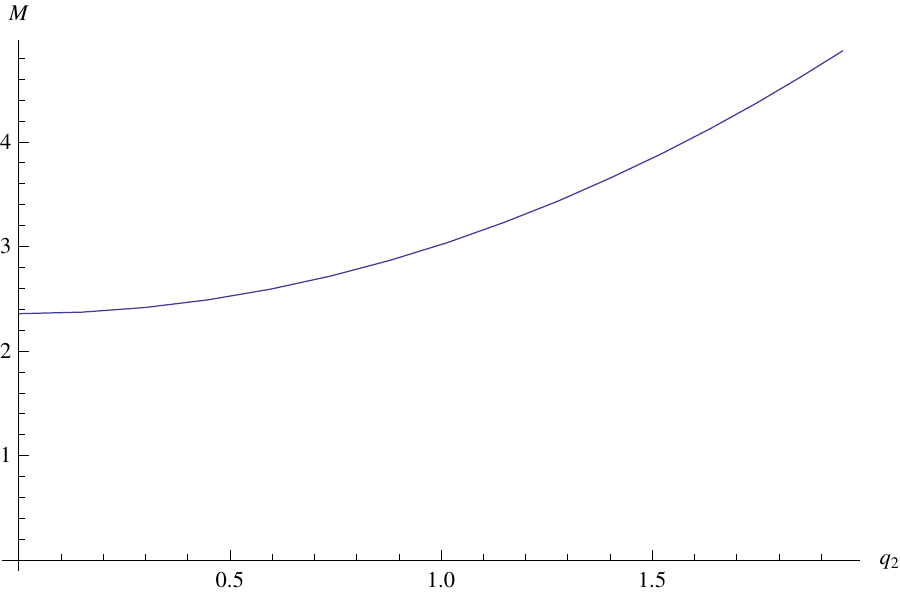}
\caption{\it Black hole for $r_0=1$ and $0<q_2<2$. The solution is
Schwarzschild-AdS when $q_2=0$.  For small $q_2$, the function
$q_1(q_2)$ is linear and $M(q_2)$ is parabolic.}
\label{bh-TMq1}
\end{figure}

\begin{figure}[ht]
\qquad\qquad\qquad\qquad
\includegraphics[width=7cm]{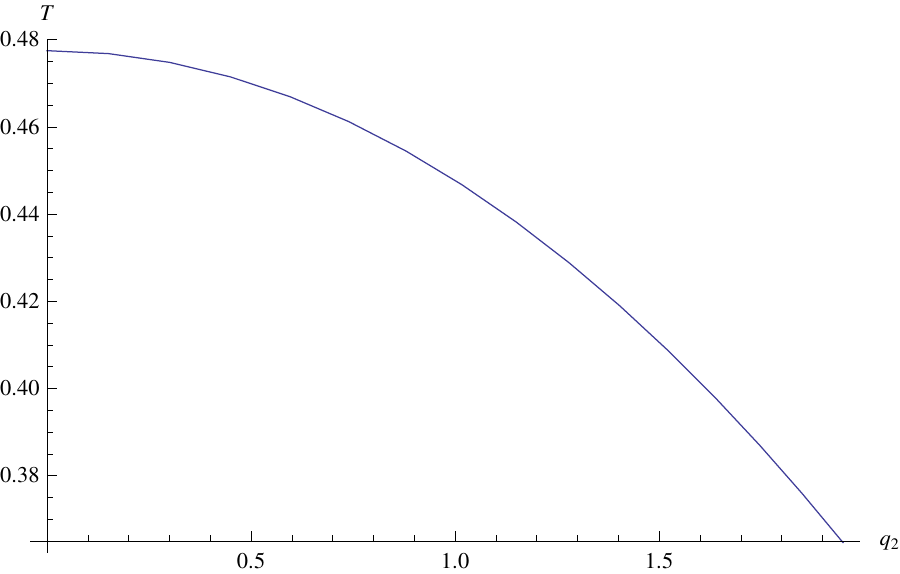}
\caption{\it Temperature of the black hole for $r_0=1$ and $0<q_2<2$.}
\label{bh-TMq1x}
\end{figure}

We can now solve numerically for a range of values for the parameter
$a_1$ and hence read off
$(q_1,q_2,M)$ and the temperature, all depending on the chosen values
of $a_1$.  In particular, when $a_1=0$, the solution becomes the usual
Schwarzschild AdS-black hole, with $r_0=1$.  Here we present the results
for $a_1$ in the range $-\ft34\le a_1\le 0$.  We can then express the
quantities $(M,q_1, T)$ as functions of $q_2$.  The results are given
in Figs~\ref{bh-TMq1} and \ref{bh-TMq1x}.

The data fitting for small $q_2$ implies that
\be
M=\ft12\pi (\ft32 + 0.420 q_2^2)\,,\qquad
q_1=-0.843 q_1\,.
\ee
Note that
\be
-\fft{2}{\pi}\fft{\partial M}{\partial q_2} =
- 0.840 q_2\,,
\ee

\begin{figure}[ht]
\ \ \ \ \  \includegraphics[width=7cm]{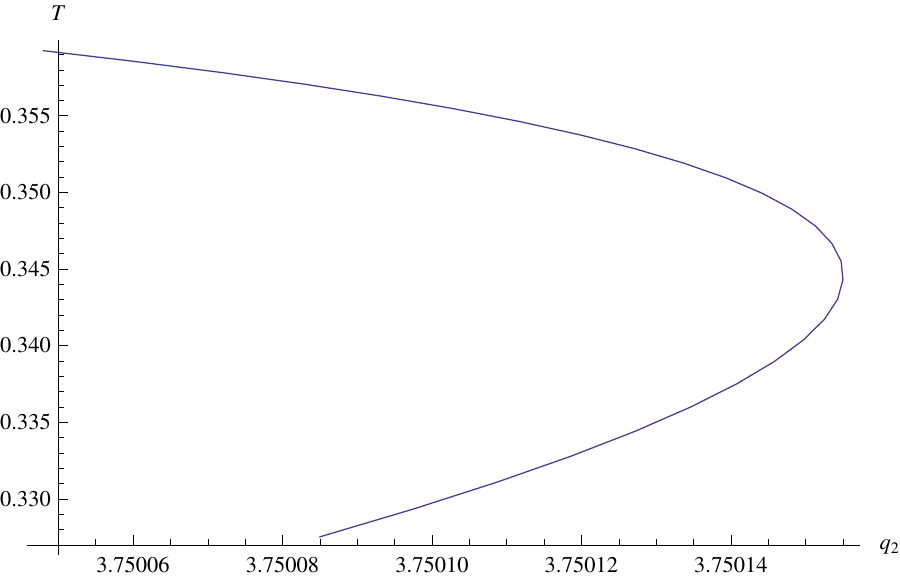}\ \ \
\includegraphics[width=7cm]{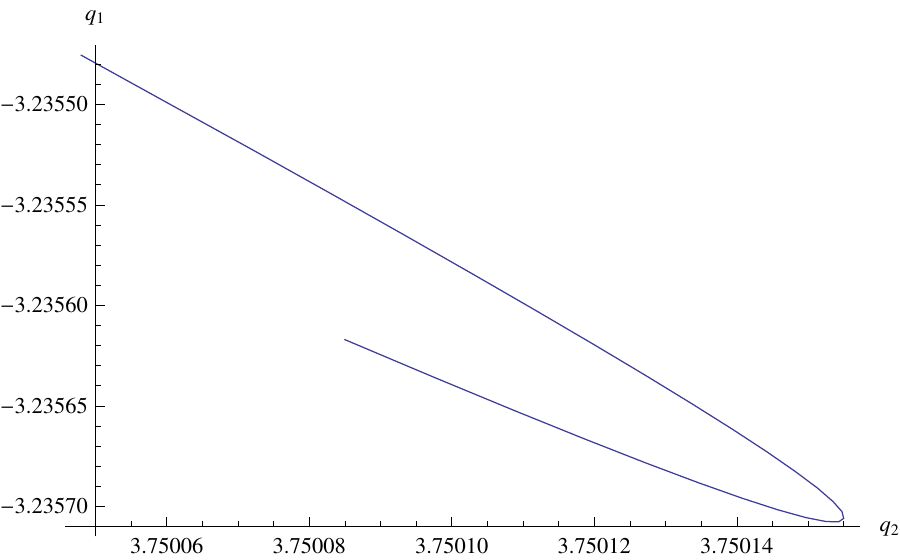}
\caption{\it Black hole for $r_0=1$ and larger $q_2$.  Note that for a
given $q_2$, there can be two values for
$q_1$ or $T$ respectively, suggesting that a phase transition can occur.}
\label{bhTq1}
\end{figure}

If we decrease $a_1$ from 0 to negative values, eventually the solution
becomes singular when $a_1$ reaches about $a_1^* \sim -20.12$. We
obtained data from $a_0=-18$ to $a_1=-20$, and the results for $T$ and
$q_1$ are plotted in Fig.~\ref{bhTq1}.

What is curious is that as $a_1$ approaches $a_1^*$, the parameters
$(q_1,q_2,m_2)$ remain finite, approaching certain fixed values.
Yet the solution becomes singular once $a_1$ passes over $a_1^*$.

\section{Conclusions}

    In this paper, we have investigated the static spherically-symmetric
solutions of the theory of a massive Proca field coupled to gravity,
in the presence of a negative cosmological constant.  The fact that the
solutions are asymptotic to anti-de Sitter, rather than Minkowski, spacetime
has a profound effect on their geometry and stability.  In the
absence of a cosmological constant, a generic static spherically-symmetric
solution of the Proca equation will take the form $\psi=A_0\sim
\alpha\, e^{-\tilde m r}/r^{n-3}+ \beta\, e^{\tilde m r}/r^{n-3}$,
and so without a fine-tuning to set $\beta=0$, the
solution will diverge exponentially at infinity.  There will in turn be
back-reaction on the metric that leads to analogous singular behaviour.
This implies that even if one finds a solution that is well behaved on the
horizon of a black hole, its evolution out to large $r$ will inevitably
pick up some component of the diverging asymptotic solution, thus implying
that it will be singular.  The asymptotic
solutions would instead be decaying and
oscillatory in $r$ if $\tilde m^2$ were negative, but then, the Proca
field would be tachyonic and so the solutions would exhibit runaway
behaviour with real exponential time dependence.

    By contrast, with the cosmological constant turned on, two
factors come into play that radically change the picture.  First of all,
the asymptotic behaviour of the Proca solutions in the AdS background
involve power-law rather than exponential dependence on $r$.  Secondly,
there is now a window of negative mass-squared values for the Proca
mass, extending in the range ${\bf m}_{\rm BF}^2 \le \tilde m^2 <0$,
where ${\bf m}^2_{\rm BF}$ is the Breitenlohner-Freedman bound given in
(\ref{BFbound}), within which the Proca field is still non-tachyonic, and
thus is not subject to exponential time-dependent runaway behaviour.
Between them, these two factors imply that perfectly well-behaved black
hole solutions exist, provided that the Proca mass-squared lies in an
appropriate range.  There also exist solitonic solutions, that extend
smoothly to an asymptotically AdS region from an origin of the radial
coordinate at $r=0$.

    We performed some numerical integrations to
demonstrate the existence of such well-behaved black  hole and solitonic
solutions, but in fact, one can see on general grounds that they must
exist.  Namely, by making a general expansion of the Proca and metric functions
in the vicinity of the horizon, in the black hole case, or of the
origin, in the solitonic case, one can first establish that well-behaved
short-distance solutions exist in each case.  Although one does not know
precisely how these join on to the solutions at large distance, which
are known only asymptotically, the fact that the general asymptotic solutions
are well-behaved means that the evolution from small to large $r$ will
necessarily be a smooth one.  In other words, there is no issue in this
asymptotically-AdS situation of needing a ``fine-tuning'' to avoid an
evolution to a singular solution at infinity, since all large-$r$ solutions
are non-singular.

    The calculation of the mass of the static spherically-symmetric
solutions can be somewhat delicate, because the way in which they approach
AdS at infinity may involve fall-offs that are slower than in a typical
AdS black hole such as Schwarzschild-AdS.  In the case of Schwarzschild-AdS,
the metric functions $h$ and $f$ in (\ref{solans}) take the form
$h=f=r^2\, \ell^{-2} + 1 - m/r^{n-3}$ in $n$ dimensions, and one might
think that any approach to pure AdS that was slower than the $1/r^{n-3}$
rate would lead to a diverging ``mass.''   We performed our calculations
of the mass using the renormalised holographic stress tensor, and
it turns out that when
the contribution from the Proca field is properly taken into account, the
result in general is perfectly finite and well defined.  We also
carried out a derivation of the first law of thermodynamics using
the techniques developed by Wald, and we found that this gives consistent
and meaningful results.

   An important feature in the first law is that
there is a contribution from the Proca field, giving a result typically
of the form
$dM= TdS + (\hbox{const})\, q_1 dq_2$, where $q_1$ and $q_2$ are the
two arbitrary coefficients in the asymptotic form of the Proca
solution.  One might think of $q_2$ as being like a ``charge'' for the
Proca field and $q_1$ as a conjugate ``potential.''  However,
since $q_2$ is not
associated with a conserved quantity it is not necessarily
clear whether it should really be thought of as a ``charge,'' or whether
instead it should be viewed as being a parameter characterising a Proca
``hair.''   In any case, the inclusion of the $q_1 dq_2$ term is necessary
in order to obtain an integrable first law.  The situation is in fact
analogous to one that was encountered for the gauged dyonic black hole
in \cite{gaugedyon}, where a careful examination of the contribution
of a scalar field in the first law showed that it gave an added
contribution, with the leading coefficients in the asymptotic
expansion of the scalar field being the extra thermodynamic variables.

\section*{Acknowledgements}

  We are grateful to Sijie Gao and Yi Pang
for useful discussions.  H-S.L. is supported in part by
NSFC grant 11305140 and SFZJED grant Y201329687.
The research of H.L. is supported in part by
NSFC grants 11175269 and 11235003. The work of C.N.P.
is supported in part by DOE grant DE-FG02-13ER42020.

\appendix

\section{Proca Solutions in AdS$_n$, and the Breitenlohner-Freedman Bound}

   Here we record some results for spherically-symmetric solutions of
the Proca equation in a pure AdS$_n$ background metric.  These allow us
to study the asymptotic form of the Proca field in our black hole and
soliton solutions, and also to give a simple derivation
of the Breitenlohner-Freedman bound for massive vector modes in $n$-dimensional
anti-de Sitter spacetime.

   From the last equation in (\ref{einstproceom}), we see that in
a pure AdS background, which has $h=f=1+r^2\, \ell^{-2}$, the Proca
potential $\psi$ for a static spherically-symmetric field $A=\psi(r) dt$
satisfies
the equation
\be
\fft1{r^{n-2}}\, (1+ r^2\, \ell^{-2})\,
 \big( r^{n-2}\, \psi'\big)'= \tilde m^2\, \psi \,.
\ee
This can be solved straightforwardly in terms of hypergeometric functions,
giving
\bea
\psi &=& \fft{q_1}{r^{(n-3-\sigma)/2}}\, F\Big(\fft{3-n-\sigma}{4},
 \fft{n-3-\sigma}{4}, \fft{2-\sigma}{2}, -\fft{\ell^2}{r^2}\Big) \nn\\
&&\qquad +
\fft{q_2}{r^{(n-3+\sigma)/2}}\, F\Big(\fft{3-n+\sigma}{4},
 \fft{n-3+\sigma}{4}, \fft{2+\sigma}{2}, -\fft{\ell^2}{r^2}\Big)\,,
\label{psisol}
\eea
where
\be
\sigma =\sqrt{4 \tilde m^2 \,\ell^2 + (n-3)^2}\,,\label{sigmasol}
\ee
and $\tilde m$ is the mass of the Proca field.
(Since our focus is principally on the large-$r$ asymptotic behaviour
of the solutions, we have presented them in the form where the hypergeometric
functions are analytic functions of $1/r^2$.)  Note that the leading-order
terms $q_1\, r^{(3+\sigma-n)/2}$ and $q_2\, r^{(3-\sigma-n)/2}$ associated
with the two independent solutions then have ``descendants'' falling off
with the additional factors of integer powers of $1/r^2$.

   The Breitenlohner-Freedman bound for the vector modes is determined by
the requirement that the parameter $\sigma$ appearing in (\ref{psisol})
should be real.  Thus from (\ref{sigmasol}) we see that the bound is
given by
\be
\tilde m^2\ge m_{\sst{BF}}^2 \equiv - \fft{(n-3)^2}{4\ell^2}\,.\label{BFdef}
\ee

   A  case of particular interest in this paper is when the
Proca mass $\tilde m$ is chosen to be given by (\ref{Procamass}), in order
to ensure that the sequence of terms in the power-series expansion of
$\psi(r)$ at large $r$ should involve inverse powers of $r$ that increase
in steps of $1/r$.  Thus we see from (\ref{BFdef}) that our
mass parameter $\tilde m$ lies within the Breitenlohner-Freedman bound, with
\be
\tilde m^2 = m_{\sst{BF}}^2 + \fft1{4\ell^2} > m_{\sst{BF}}^2\,.
\label{within}
\ee
We then study a more extended range of values for $\tilde m$.

  Note that in the limit where the cosmological constant goes to zero
(i.e. $\ell\rightarrow\infty$), the Proca solution (\ref{psisol})
becomes
\be
\psi = \fft{\alpha\, e^{-\tilde m r}}{r} +
  \fft{\beta\, e^{\tilde m r}}{r}\,.\label{psimink}
\ee
Unlike the AdS case, where both solutions of the Proca equation can be
well-behaved at infinity, in an asymptotically-Minkowski background one
of the solutions always diverges exponentially.

\end{document}